\documentclass[%
reprint,
superscriptaddress,
showkeys,showpacs,preprintnumbers,
amsmath,amssymb,
aps,
prb,
]{revtex4-2}

\usepackage{graphicx}
\usepackage{dcolumn}
\usepackage{bm}

\usepackage{upgreek}
\usepackage{amsmath}

\begin{document}


\title{Phase-change nonlocal metasurfaces for dynamic wavefront manipulation}

\author{Tingting Liu}
\affiliation{Institute for Advanced Study, Nanchang University, Nanchang 330031, China}
\affiliation{Jiangxi Key Laboratory for Microscale Interdisciplinary Study, Nanchang University, Nanchang 330031, China}

\author{Dandan Zhang}
\affiliation{School of Physics and Materials Science, Nanchang University, Nanchang 330031, China}

\author{Wenxing Liu}
\affiliation{School of Physics and Materials Science, Nanchang University, Nanchang 330031, China}

\author{Tianbao Yu}
\affiliation{School of Physics and Materials Science, Nanchang University, Nanchang 330031, China}

\author{Feng Wu}
\affiliation{School of Optoelectronic Engineering, Guangdong Polytechnic Normal University, Guangzhou 510665, China}

\author{Shuyuan Xiao}
\email{syxiao@ncu.edu.cn}
\affiliation{Institute for Advanced Study, Nanchang University, Nanchang 330031, China}
\affiliation{Jiangxi Key Laboratory for Microscale Interdisciplinary Study, Nanchang University, Nanchang 330031, China}

\author{Lujun Huang}
\email{ljhuang@phy.ecnu.edu.cn}
\affiliation{School of Physics and Electronic Sciences, East China Normal University, Shanghai 200241, China}

\author{Andrey E. Miroshnichenko}
\email{andrey.miroshnichenko@unsw.edu.au}
\affiliation{School of Engineering and Technology, University of New South Wales, Canberra, ACT 2612, Australia}

\begin{abstract}
Recent advances in nonlocal metasurfaces have enabled unprecedented success in shaping the wavefront of light with spectral selectivity, offering new solutions for many emerging nanophotonics applications. The ability to tune both the spectral and spatial properties of such a novel class of metasurfaces is highly desirable, but the dynamic nonvolatile control remains elusive. Here, we demonstrate active narrowband wavefront manipulation by harnessing quasi-bound states in the continuum (quasi-BICs) in phase-change nonlocal metasurfaces. The proof-of-principle metasurfaces made of Sb$_2$S$_3$ allow for nonvolatile, reversible, and tunable spectral control over wavefront and switchable spatial response at a given wavelength. The design principle mainly builds upon the combination of the geometry phase of quasi-BICs and the dynamic tunability of phase-change meta-atoms to tailor the spatial response of light at distinct resonant wavelengths. By tuning the crystallization level of Sb$_2$S$_3$ meta-atoms, the dynamic nonlocal wavefront-shaping functionalities of beam steering, 1D, and 2D focusing are achieved. Furthermore, we demonstrate tunable holographic imaging with active spectral selectivity using our phase-change nonlocal metasurface. This work represents a critical advance towards developing integrated dynamic nonlocal metasurface for future augmented and virtual reality wearables.

\end{abstract}

\maketitle


\section{\label{sec1}Introduction}

The past few decades have witnessed substantial progress of metasurfaces in control over the light field through subwavelength elements, as they can be designed with a variety of geometry and materials to generate a tailored optical response in terms of amplitude, phase, polarization, frequency, and other degrees of freedom \cite{Kildishev2013,Yu2014,Ding2017,So2023,Arbabi2022}. Especially with their ability to impart a specific phase profile using designed meta-atoms for wavefront control, metasurfaces have provided an elegant and fashionable strategy to miniaturizing and flattening various optical systems \cite{yu2011light,sun2012gradient,Chen2018,Lin2019,li2020three,Georgi2021,Guo2021,Luo2022,ren2020complex,Li2022a,Li2023}. The conventional metasurfaces are designed to imprint independent and spatially variant phases onto incident light waves, and these local devices would shape the wavefront over a broad range of wavelengths due to the modulation as a function of position. However, in many emerging imaging, sensing, and display applications such as augmented reality (AR) and virtual reality (VR) systems, the new functionalities that require highly localized energy and high spectral control over the wavefront are inaccessible by the local metasurfaces. Most recently, several works have employed the concept of bound states in the continuum (BICs) to realize the nonlocal wavefront-shaping metasurfaces with large quality factor ($Q$-factor) and spectral selectivity \cite{Overvig2020,Sang2022,overvig2023demonstration,Huang2023,Xu2023}. Based on the analysis of symmetry perturbations to the BICs, the general method for independently controlling local and nonlocal interaction of the metasurface supporting quasi-BICs is developed to shape wavefront by spatially varying the polarization properties of quasi-BICs only on resonance, leaving the non-resonant light unchanged. The nanostructures such as wavefront-selective metasurface\cite{Overvig2021} and spectrally selective metalenses \cite{Overvig2022,Chen2022,Malek2022}, have been theoretically and experimentally demonstrated, opening new pathways for future 3D display platforms that complement those of conventional metasurfaces.

Despite the exciting advances in nonlocal metasurfaces, the ability to realize flat optical devices with controllable spectral and spatial response remains a significant ongoing challenge for practical applications. In previous works, researchers have realized the control of multiple quasi-BICs by adding successive symmetry-breaking perturbations to a single nonlocal metasurface, allowing for wavefront manipulation at several wavelengths \cite{Overvig2020a,Zhou2023}. In principle, up to four distinct functions from four quasi-BICs may be realized on a single metasurface though at the cost of denser patterning and increased cross-talk. Besides the single-layer structures, multiple nonlocal wavefront-shaping metasurfaces can also be stacked to mold the optical wavefront distinctively at multiple wavelengths \cite{Malek2022}. However, these approaches require delicate design and precise fabrication techniques. Moreover, such passive metasurfaces are designed to be static with fixed control over spatial and spectral response once they are fabricated. Inspired by the progress of tunable and reconfigurable nanophotonics \cite{Ding2019, Xiao2020,Kim2022,Abdelraouf2022}, recent efforts have explored active tunability and reconfigurability of wavefront-shaping functionalities in nonlocal metasurfaces. For example, by leveraging tunable thermo-optic coefficient of titanium dioxide (TiO$_2$) material, the resonant wavelengths of quasi-BICs in TiO$_2$ nonlocal metasurface are thermally shifted through refractive-index tuning, and then the thermally switchable functionalities have been demonstrated in the metasurface \cite{malek2023thermally}. In parallel, the spatial and spectral reconfigurability of nonlocal metasurfaces has also been explored in a mechanically stretchable deflector consisting of silicon pillars embedded in a stretchable polymer, where resonant wavelength and deflection angle can be simultaneously tunable via mechanical strain\cite{Malek2020}. The thermally tunable and stretchable metasurfaces show the advantages of simple design, fabrication, and accessible tuning approaches. Nevertheless, once the thermal control or mechanical strain vanishes, the spectral and spatial responses of the metasurfaces recover to their initial states.

This work employs an emerging phase-change material, antimony trisulfide (Sb$_2$S$_3$), in nonlocal metasurfaces supporting quasi-BICs for dynamic spatial and spectral control. By tuning the crystallization level of Sb$_2$S$_3$, wavefront shaping can be realized at different wavelengths, and the optical functionalities at specific wavelengths can be switched on to off. Compared with the previous tuning mechanism, the approach based on phase-change nonlocal metasurface shows that multiple functionalities can be integrated into a single metasurface with active, nonvolatile, and reversible tunability but without complex patterning, affording more degrees of freedom for light manipulations. As a proof-of-concept, we design the Sb$_2$S$_3$ metasurface supporting the quasi-BIC that is evolved from the symmetry-protected BIC by the dimerization perturbation. The quasi-BIC obeys the selection rules, and the system is capable of shaping wavefront only across the narrow spectral bandwidth of resonance using the geometric phase. By tuning the crystallization level of Sb$_2$S$_3$ meta-atoms, the dynamic spectral and spatial control for functionalities such as beam steering, cylindrical lensing, and radial lensing is demonstrated. Furthermore, the dynamic holographic imaging at distinct wavelengths is demonstrated, showing the great potential of the phase-change nonlocal metasurfaces in optical information storage and encryption. By introducing the meta-atoms made of the phase-change materials into the nonlocal wavefront-shaping metasurface, the active control of wavefront manipulation at narrow spectral bandwidth of the quasi-BIC resonances is realized with the advantage of multistate, reversible, and nonvolatile tunability, which is the first time in active nonlocal metasurfaces. It may lead to new device concepts and find great potential in optical imaging, sensing, and display applications.

\section{\label{sec2}Design Principle}

Figure 1 is a conceptual illustration of phase-change nonlocal metasurface enabling dynamic spatial control at distinct resonant wavelengths by tuning the crystallization level of Sb$_2$S$_3$. In the proposed metasurface, the wavefront manipulation stems from the geometric phase imparted onto the converted circularly polarized (CP) light of transmission mode on resonance. The geometric phase only relies on the polarization properties of quasi-BIC from symmetry breaking. Meanwhile, the quasi-BIC resonance of the metasurface strongly depends on the refractive index of Sb$_2$S$_3$ meta-atoms, which is tunable by adjusting the crystallization state of Sb$_2$S$_3$ through controlling the external stimulus. Under the operating principle, dynamic wavefront manipulation can be realized at the tunable spectral resonance in a specific symmetry-broken quasi-BIC nanostructure. As shown in Fig. 1, the wavefront would be shaped at the resonant wavelength when the broadband CP light illuminates onto the quasi-BIC metasurface with a designed phase profile. As Sb$_2$S$_3$ metasurface experiences phase transition from the amorphous (a-Sb$_2$S$_3$) to crystalline states (c-Sb$_2$S$_3$), the spectral response is shifted from $\lambda_a$ to $\lambda_c$, tuning the centered wavelength of spatial control. Especially, Sb$_2$S$_3$ material possesses multilevel nonvolatile tunable optical constants and the ultrafast reversible switching ability between the stable states. Benefiting from these advantages, Sb$_2$S$_3$ metasurface can realize dynamic, reversible, and nonvolatile wavefront manipulation at multiple distinct wavelengths during the phase transition process. 

\begin{figure*}[htbp]
	\centering
	\includegraphics[width=\linewidth]{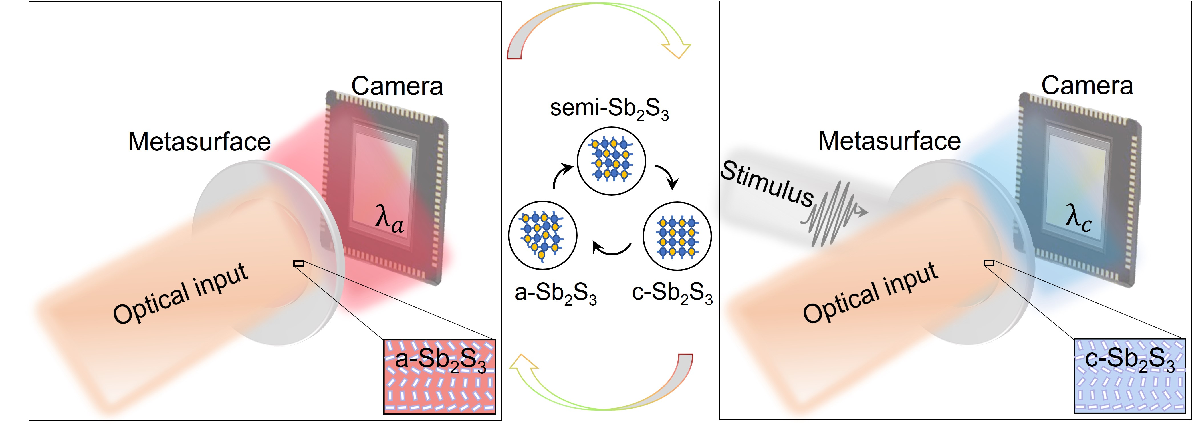}
	\caption{The schematic diagram of dynamic wavefront manipulation at distinct wavelengths enabled by the proposed phase-change nonlocal metasurfaces. Using the Sb$_2$S$_3$ metasurface, the wavefront of optical input is manipulated at the resonant wavelength of quasi-BIC. Through external stimulus of light or heat, Sb$_2$S$_3$ meta-atom is phase changed between the amorphous (a-Sb$_2$S$_3$) and crystalline states (c-Sb$_2$S$_3$), leading to the resonant wavelength shifting between $\lambda_a$ and $\lambda_c$. Inset in the middle is a generic scheme of the atomic distributions of the three crystallinity states including a-Sb$_2$S$_3$, the intermediate state referred to semi-Sb$_2$S$_3$, and c-Sb$_2$S$_3$, which are nonvolatile stable states that can be reversibly switched during the phase transition process. The resultant multi-state spectral shift of quasi-BIC resonance gives rise to a tunable center wavelength for the dynamic spatial response.}
	\label{fig1}
\end{figure*}

The proposed phase-change metasurface can manipulate the optical wavefront by choosing a specific wavelength within a broad regime, which is different from the previous nonlocal metasurfaces with only several target working wavelengths. This can be attributed to multilevel spectral control over wavefront in Sb$_2$S$_3$ metasurfaces. The proposed metasurface has more freedoms and higher security for information encryption. For example, the information decoding at a selected wavelength requires an appropriate stimulus to transform Sb$_2$S$_3$ at a given crystallization level, and misleading or useless information will be obtained either for other wavelengths or for other improper crystallisation levels. More importantly, it offers an important solution for wearables and AR devices. For example, it can not only satisfy the need to shape optical wavefront exclusively at selected wavelengths and leave the rest of the spectrum transparent but also bring the flexibility of adaptive control with adjusting the working wavelength of wavefront manipulation. The proposed metasurfaces offer a prototype for potential applications such as information storage and encryption, AR and display technology.

To present the operation mechanism of the dynamic control of the spectral and spatial response, we first elucidate the evolution of the quasi-BIC resonance in the metasurface. BICs can be characterized as nonradiating resonant modes in an open system but without coupling to the radiating channels propagating outside the system \cite{Hsu2016,Sadreev2021,Huang2023a}. Here we focus on the symmetry-protected BICs developed by symmetry-restricted coupling to free space. To be excitable from free space light for real applications, the leaky quasi-BICs with finite $Q$-factors are induced by a symmetry-breaking perturbation. An alternative approach is the dimerization perturbation to double the period along the real-space dimension and halve the first Brillouin zone in momentum space. In Fig. 2(a), the simple structure comprising a rectangular aperture etched in a dielectric thin film is adopted for demonstration. Figure 2(b) shows the corresponding dimerized meta-atom with a two-fold symmetry. As compared in Figs. 2(c) and 2(d), the period-doubling perturbation effectively folds the band structure in the $k$-space, such that the bound mode that was under the light line and not radiated to free space is folded into the continuum at the $\Gamma$ point that can be excitable by free space light under normal incidence. In this evolution, the optical lifetime of such quasi-BIC can be controlled by the perturbation strength ($\delta$) with $Q$-factors varying inversely with $\delta$ as $Q\propto\delta^{-2}$, and the polarization properties are governed by selection rules specifying whether excitation of quasi-BIC is forbidden or allowed according to the space group of the perturbed symmetry \cite{Koshelev2018,Li2019,Wang2020,Overvig2020,Xiao2022}. Thus it is possible to attain the desired $Q$-factor by adjusting the perturbation strength and realize the wavefront control by manipulating the polarization properties of quasi-BIC in a designed metasurface.

\begin{figure}[htbp]
	\centering
	\includegraphics
	[width=\linewidth]{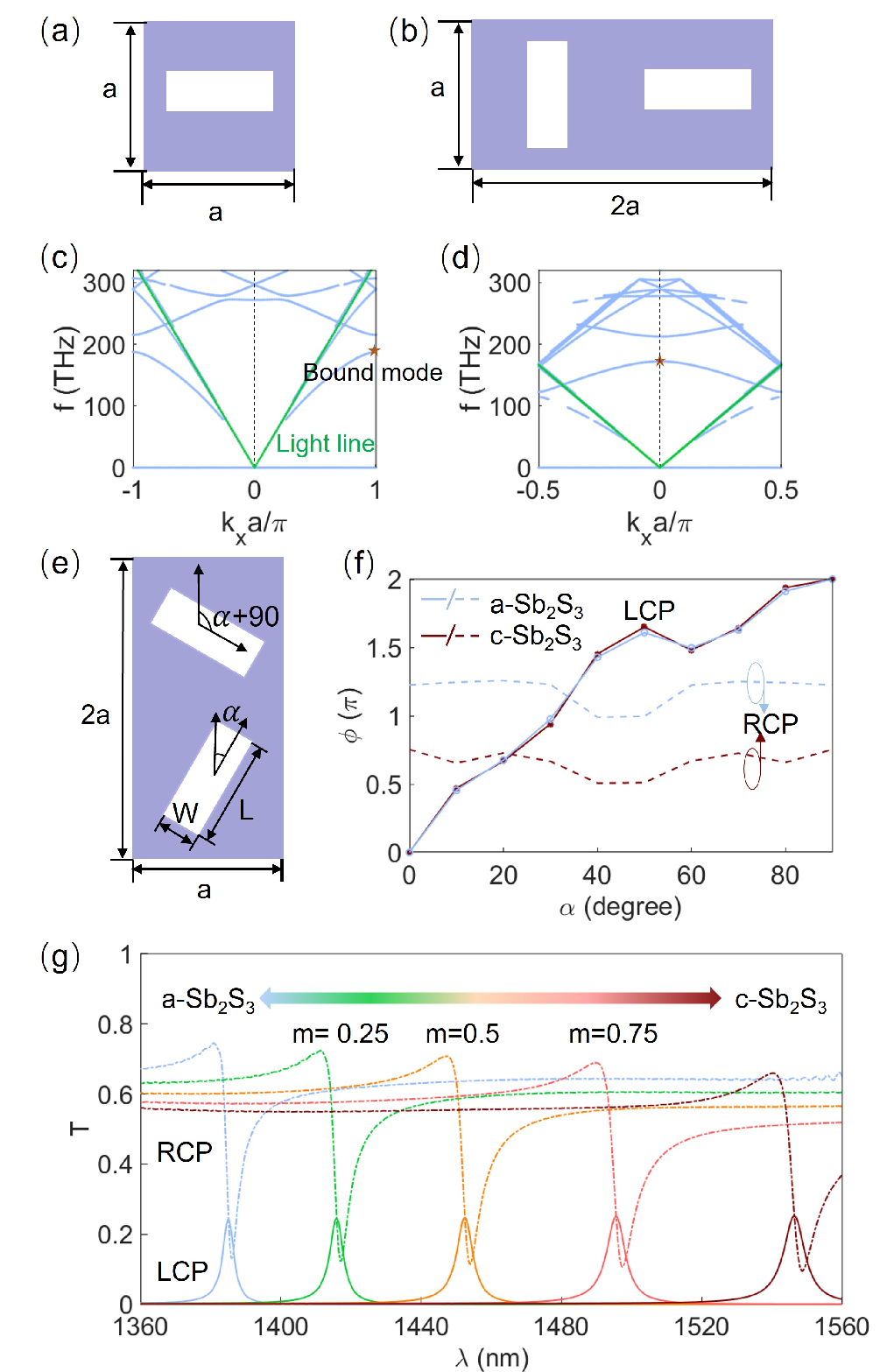}
	\caption{Design of phase-change nonlocal metasurface. (a)-(d) The evolution of quasi-BIC from the symmetry-protected BIC via the dimerization perturbation. (a) The unperturbed meta-atom with a rectangular aperture etched in a silicon thin film. (b) The dimerized meta-atom. The period is $a=450$ nm, the etched hole length is $L=D+\delta$ and the width $W=D-\delta$ with $\delta=100$ nm and $D=225$ nm. The silicon film with a thickness of 150 nm is used. (c) and (d) The calculated band structures for the unperturbed meta-atoms in (a) and for the dimerized meta-atoms in (b), respectively. The green curve depicts the light line, and the star mark indicates the mode of interest.  (e)-(g) Design of diatomic phase-change resonators and their optical properties. (e) Top view of the designed meta-atoms showing its geometrical parameters of $a=450$ nm, $L=325$ nm, $W=125$ nm, and varying rotation angles $\alpha$. (f) The converted LCP and unconverted RCP components' simulated phase in transmission for the a-Sb$_2$S$_3$ and c-Sb$_2$S$_3$ metasurfaces when the RCP is normally incident. (g) Simulated transmission spectra of the converted LCP and unconverted RCP component in transmission at different crystallization levels when the RCP is normally incident on the metasurface. }
	\label{fig2}
\end{figure}

Then we consider the dimerized structure consisting of two rectangular holes in Sb$_2$S$_3$ thin film with a thickness of 150 nm on the silica substrate, categorized as $p2$ space group, as shown in Fig. 2(e). To concentrate on the dynamic wavefront control, the $Q$-factor engineering involved is not taken into consideration in the current work, and we set the constant symmetry-breaking strength of the structure as $\delta=100$ nm with the hole length as $L=D+\delta$ and width $W=D-\delta$. Independently of this optical localisation and lifetime, the polarization angle $\phi$ of the quasi-BIC coupling to free space light can be controlled by the in-plane rotation angle $\alpha$, with a linear approximation of $\phi\approx2\alpha$, based on the selection rules \cite{Overvig2020}. This is derived from the parent-child symmetry relationship where the $pmm$ parent space with $\alpha=0^{\circ}$ allows for mode coupling to $x$ polarization with $\phi=0^{\circ}$ and the $pmg$ parent space with $\alpha=45^{\circ}$ yields coupling to $y$ polarization with $\phi=90^{\circ}$. Like the conventional phase-gradient metasurface, the $p2$ space group system imparts the geometric phase to the converted CP light in transmission. Specifically, when the CP light is incident onto the metasurface, only the linear component in the polarization angle of $\phi_1 \approx 2\alpha$ completely coupling to the quasi-BIC, and subsequently, the linear polarized light on resonance is coupled out into the free space with decomposed CP component showing phase as $\phi_2\approx2\alpha$ for the converted part. Hence, the total geometric phase of the quasi-BIC in this process is derived as $\Phi=\phi_1+\phi_2\approx4\alpha$ for the converted CP component, while the phase remains unchanged for the unconverted component due to the opposite signs canceling each other when transforming the linear and CP components. Such relations can be illustrated by Fig. 2(f), where the geometric phase of quasi-BIC on resonance depends on the rotation angle in the Sb$_2$S$_3$ meta-atom. When the right-handed circularly polarized (RCP) light illuminates on the metasurfaces, the geometric phase of the transmitted light on quasi-BIC resonance shows an approximate relation of $4\alpha$ for the converted component of LCP with a full phase coverage of $0\sim2\pi$, whereas the phase for the unconverted component of RCP is nearly unchanged.  

Compared with the common phase-change material Ge$_2$Sb$_2$Te$_5$, Sb$_2$S$_3$ has a wide bandgap of $1.7\sim2.05$ eV, showing much lower, even negligible absorption in the near-infrared regime of interest \cite{lu2021reversible,lu2022reconfigurable,Moitra2022}. The refractive index of Sb$_2$S$_3$ can be adjusted from 2.8 to 3.5 in this regime (shown in Figure S1). Via appropriate stimulus, including optical, thermal, and electric excitations, Sb$_2$S$_3$ can be switched reversibly and quickly between the amorphous (a-Sb$_2$S$_3$) and crystalline states (c-Sb$_2$S$_3$) or to be an intermediate state (semi-Sb$_2$S$_3$). The dielectric constant at different crystallization levels of Sb$_2$S$_3$ can be derived by the Lorenz-Lorentz relation as \cite{Tian2019,meng2021high,Liu2022} 
\begin{equation}
\frac{\varepsilon_{\text{eff}}-1}{\varepsilon_{\text{eff}}+2}
=m\frac{\varepsilon_{c}-1}{\varepsilon_{c}+2}
+(1-m)\frac{\varepsilon_{a}-1}{\varepsilon_{a}+2},
\end{equation}
where $m$ represents the crystalline fraction ranging from 0 to 1 for the multistate intermediating between a-Sb$_2$S$_3$ and c-Sb$_2$S$_3$, $\varepsilon_{a}$ and $\varepsilon_{c}$ are the permittivity of a-Sb$_2$S$_3$ and c-Sb$_2$S$_3$, respectively. Because of the tunable optical constants of meta-atoms, the spectral response of quasi-BIC in Sb$_2$S$_3$ metasurface will vary during these phase transitions. The numerical calculations are performed with the finite-difference-time-domain (FDTD) method via the ANSYS Lumerical Suit. As shown in Fig. 2(g), the transmission spectra show a shift of resonance wavelength within a broad band from $\sim1380$ nm to $\sim1550$ nm between a-Sb$_2$S$_3$ and c-Sb$_2$S$_3$ metasurfaces. For the converted light component of interest, namely, LCP, the peak transmission efficiencies of these spectra are always around the maximum theoretical limit of 0.25 in this four-port system \cite{Malek2020}. In contrast with the spectral shift, the geometric phase of quasi-BIC on resonance mainly relying on the rotation angles is almost unchanged, as verified by that of the a-Sb$_2$S$_3$ and c-Sb$_2$S$_3$ metasurface in Fig. 2(f). A designer phase profile can be realized at tunable wavelengths of quasi-BIC resonances. Of course, the wavefront shaping ability in metasurfaces will be remarkably weakened at the non-resonance wavelengths due to the dramatically reduced transmission efficiency. This makes it possible to switch on and off the spatial response at a specific wavelength by transforming the resonant wavelength into non-resonant via varying the crystallization levels of Sb$_2$S$_3$. As a result, dynamic wavefront-shaping metasurface supporting a quasi-BIC can be implemented by a rational synthesis of multistate switchable spectral response in the Sb$_2$S$_3$ metasurface with a spatially varying geometric phase profile. Note that Sb$_2$S$_3$ is patterned into nanostructures to construct the nonlocal metasurface instead of the spacer layer in earlier works for active devices \cite{Lee2017,Zhou2020,Mao2020,Liu2021}. The nanofabrication techniques have been explored and proven to be feasible in previous experimental works \cite{lu2021reversible,Moitra2022}, and the fabrication process for the proposed Sb$_2$S$_3$ metasurface is illustrated in Figure S2.

\section{\label{sec3}Results and discussion}

\begin{figure}[htbp]
	\centering
	\includegraphics
	[width=\linewidth]
	{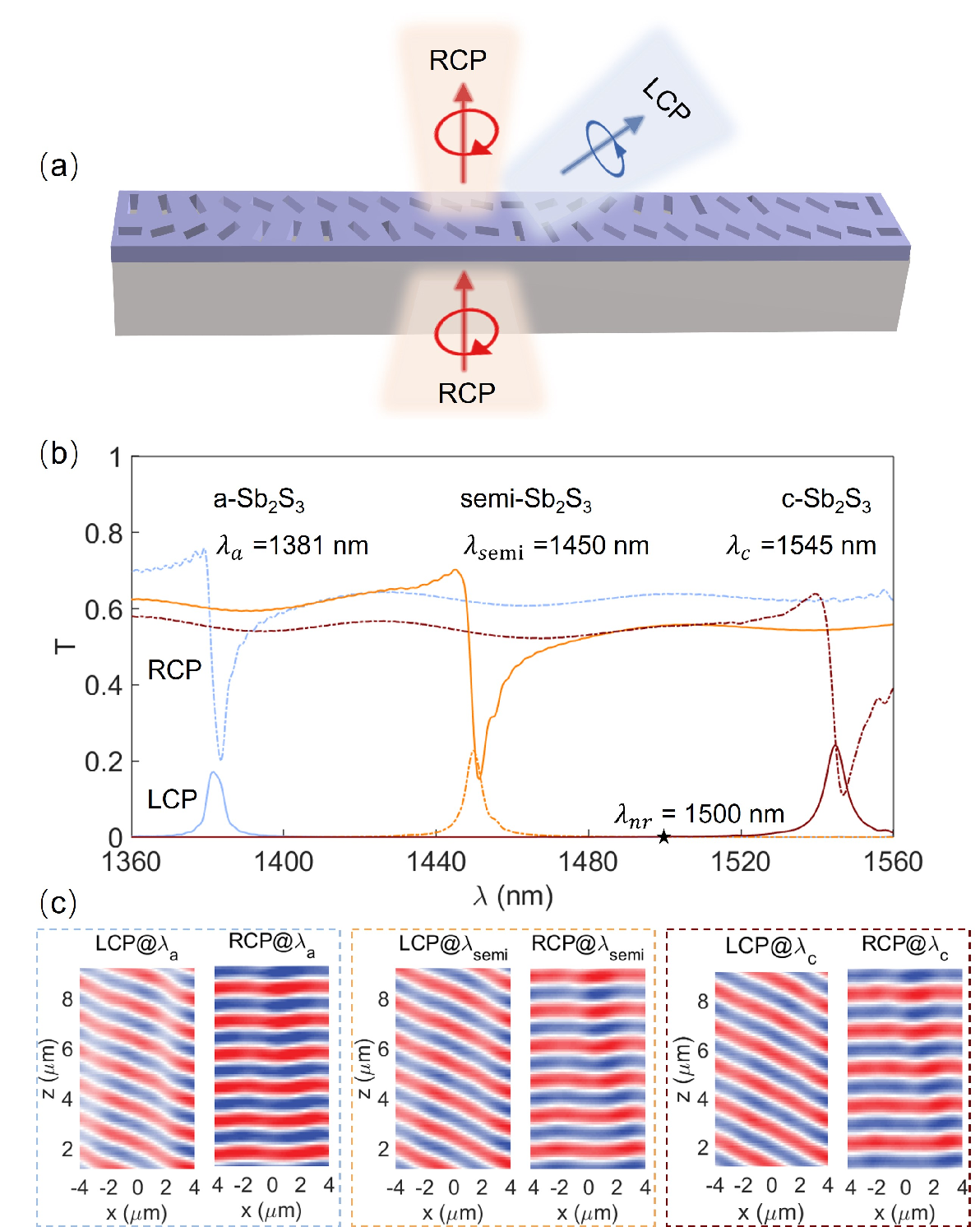}
	\caption{Design of phase-change nonlocal metasurface for dynamic beam steering at distinct wavelengths. (a) The schematic diagram of a super-unit cell of the dynamic beam deflector consisting of 25 dimerized meta-atoms with spatially varying rotation angles $\alpha$ from 0 to $\pi$. When the broadband RCP light normally illuminates onto the metasurface, the light with converted handedness is deflected in a narrow band around the resonance. (b) Transmission spectra of the nonlocal metasurface at different crystallization states of a-Sb$_2$S$_3$, semi-Sb$_2$S$_3$, and c-Sb$_2$S$_3$ for RCP and LCP component under normal RCP incidence. (c) Electric filed profiles for RCP and LCP component at the corresponding resonant wavelength of a-Sb$_2$S$_3$, semi-Sb$_2$S$_3$ and, c-Sb$_2$S$_3$ metasurface, showing the active beam deflection at different wavelengths of quasi-BIC resonances for converted LCP component.}
	\label{fig3}
\end{figure}

Using the dimerized Sb$_2$S$_3$ meta-atoms with different rotation angles for specific phase profiles, tunable spectral control over wavefront and switchable spatial response at a selected wavelength can be realized, creating active metadevices such as beam deflectors and lenses. For demonstration, we start from a common wavefront-shaping function of anomalous refraction for beam steering. According to the generalized Snell’s law \cite{yu2011light,sun2012gradient}, the nonlocal wavefront-shaping metasurface with a gradient distribution of rotation angle is devised, as shown in Fig. 3(a). The Sb$_2$S$_3$ meta-atoms are spatially arranged for one period of the metasurface with varying $\alpha$ at each position to linearly grade the corresponding geometric phase along the direction orthogonal to the dimerization, but with constant aperture dimensions and lattice constants. When RCP light is normally incident from the substrate side of the metasurface, the transmission spectra for the converted LCP and unconverted RCP component are presented in Fig. 3(b). As the crystallization fraction $m$ of Sb$_2$S$_3$ material changes from 0 to 1, namely a-Sb$_2$S$_3$ to c-Sb$_2$S$_3$, the resonance wavelength of quasi-BIC shows a multi-level redshift across a wavelength regime of $\sim160$ nm. Here, we select the intermediate state with $m=0.5$ as the semicrystalline state of Sb$_2$S$_3$ (semi-Sb$_2$S$_3$) to demonstrate the multilevel control.

As a consequence of the multilevel tunable wavelength of quasi-BIC resonance, the metasurface is able to perform wavefront shaping at different wavelengths. As shown in Fig. 3(c), the converted LCP component is refracted to $14^\circ$ at resonance wavelength of 1381 nm for a-Sb$_2$S$_3$, $15^\circ$ at 1450 nm for semi-Sb$_2$S$_3$, and $16^\circ$ at 1545 nm for c-Sb$_2$S$_3$, respectively. At these corresponding resonance wavelengths, $\sim$20$\%$ of the light is removed from the incident beam and redirected into the second-order diffracted beam\cite{Overvig2020a}. Meanwhile, no deflection for the unconverted RCP component can be observed in Fig. 3(c) due to the constant phase profiles. Unlike the conventional phase-gradient metasurfaces, the designed metasurface using resonant meta-atoms shapes the wavefront of the converted LCP component only at resonance. Such nonlocal wavefront-shaping ability can be verified by comparing the field distributions of an arbitrary non-resonant wavelength at 1500 nm (shown in Figure S3) and the respective resonant wavelengths. The phase-change metasurface demonstrates the dynamic beam deflection for the converted LCP transmission light at distinct resonant wavelengths while remaining unchanged for unconverted RCP or non-resonance.

Next we consider simple cylindrical metalens for dynamic narrowband focusing, as shown in Fig. 4(a). Under the normally incident RCP light, the transmitted LCP light is focused to a specific distance at tunable wavelengths by varying crystallization state of Sb$_2$S$_3$. The typical phase profile of the metalens with focal length $f$ is given by\cite{Chen2012,Khorasaninejad2016,Zhao2021,Li2022}
\begin{equation}
	\phi_r=\frac{2\pi}{\lambda_0}(\sqrt{r^2+f^2}-f),
	\label{eq1}
\end{equation}
where $r=\sqrt{x^2+y^2}$ is the distance from arbitrary position $(x, y)$, and $\lambda_0$ is the target working wavelength. Here, the one-dimensional metalens along $x$ direction for space light focusing is designed, with the phase profile shown in Fig. 4(b). The focal length of the metalens is 100 nm, and the working wavelength is set as the same as the resonant wavelength for a-Sb$_2$S$_3$ metasurface. The spatially varying phase profile is coded by Sb$_2$S$_3$ meta-atoms with different rotation angles $\alpha$ at corresponding coordinate, according to the nearly linear dependence of the geometric phase on the rotation angle of $\Phi \approx 4\alpha$. 

\begin{figure}
	\centering
	\includegraphics
	[width=\linewidth]
	{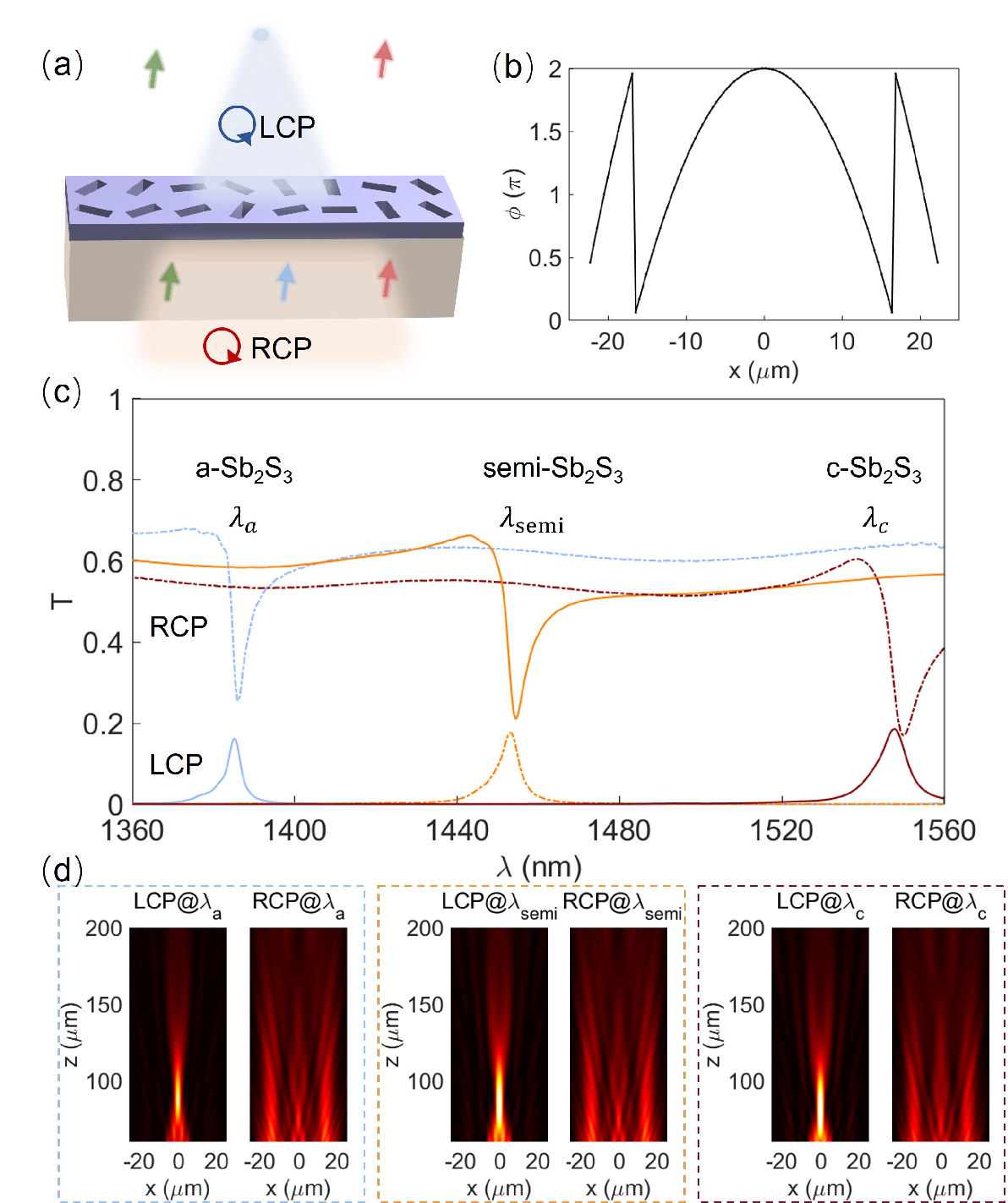}	
	\caption{Design of phase-change metasurface for dynamic cylindrical metalens at distinct wavelengths. (a) Schematic diagram of the dynamic narrowband focusing along $x$ direction. When the broadband RCP light normally illuminates onto the metasurface, the light with converted handedness is focused in a narrow band. (b) The phase profile of the 1D focuses on the converted LCP. (c) Transmission spectra of the nonlocal metasurface at different crystallization states. (d) Far-field intensity distributions near the designed focal spot of 100 nm for RCP and LCP component at the corresponding resonant wavelength of a-Sb$_2$S$_3$, semi-Sb$_2$S$_3$, and c-Sb$_2$S$_3$ metasurface, showing the space light focusing at different wavelengths of quasi-BIC resonances for converted LCP component.}
	\label{fig4}
\end{figure}

In Fig. 4(c), the transmission spectra of the cylindrical metalens demonstrate that the converted LCP light shows a peak at resonant wavelength, and the resonance peak can be shifted from $\sim$1380 nm to $\sim1550$ nm by varying the crystallization state of phase-change meta-atoms from a-Sb$_2$S$_3$ to c-Sb$_2$S$_3$. Similarly, the resonant wavelength shift for different Sb$_2$S$_3$ contributes to the tunable spatial and spectral response. Fig. 4(d) illustrates the normalized intensity of the converted LCP light fields at the corresponding resonant wavelength of a-Sb$_2$S$_3$, semi-Sb$_2$S$_3$, and c-Sb$_2$S$_3$, respectively. Since the geometric phase is fundamentally robust and emerges in symmetry breaking, the light fields at these wavelengths are all focused, implying the ability of spatial control at multiwavelength within a broad spectral range. On the other hand, for a selected wavelength, the focusing effect can be switched on-to-off by tuning the resonant position of quasi-BIC. The focusing intensity of fields shows an observable increase, accompanied by the increasing transmission peak when the meta-atoms phase changed from a-Sb$_2$S$_3$ to c-Sb$_2$S$_3$. This can be explained by the fact that the relatively larger refractive index of c-Sb$_2$S$_3$ and negligible loss account for a stronger localized light field and, thus, stronger spatial control ability. In contrast with the space light focusing of the converted LCP component, the wavefront shaping is not performed for the unconverted RCP component in Fig. 4(d) and the non-resonant wavelengths, for example, at $\lambda=1500$ nm (shown in Figure S4). Only for the converted LCP on resonance, the designed active cylindrical metalens exhibit dynamic light focusing in the narrowband of quasi-BIC resonances. 

We proceed to the more complex two-dimensional (2D) spatial control at distinct resonant wavelengths of quasi-BICs. The phase-change metasurface for a radial lens is designed in a simple-to-implement way, which focuses the transmitted LCP component into a specific spot in space, as shown in Fig. 5(a). The spatial phase profile from Eq. (1) is presented in Fig. 5(b) with a 2D distribution dependent on the coordinate $x$ and $y$, different from the 1D cylindrical lens in Fig. 4(b). But in a similar way, the light focusing with focal spots imaging at the focal plane only occurs when the incident wavelengths tune around the quasi-BIC resonance. As the geometric phase of $\Phi \approx4\alpha$ is imparted onto the transmitted LCP component on quasi-BIC resonances, the phase profile can be encoded by the 2D spatial distribution of Sb$_2$S$_3$ meta-atoms with corresponding rotation angle $\alpha$. Here, a nonlocal radial metalens with a diameter of 30 $\upmu$m is designed and it has an initial working wavelength centered at the resonant wavelength for a-Sb$_2$S$_3$ metasurface.

\begin{figure}[htbp]
	\centering
	\includegraphics[width=\linewidth]{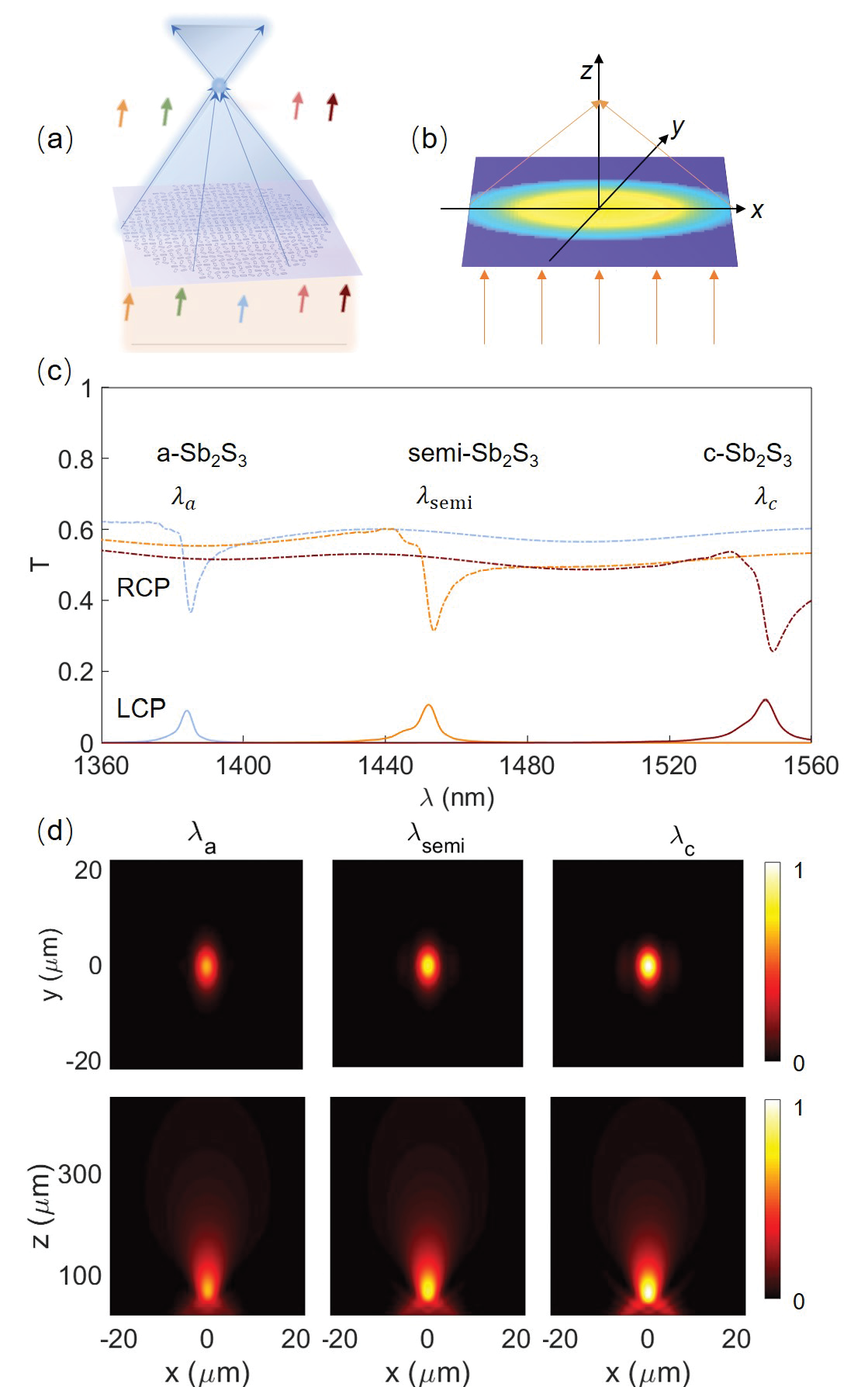}
	\caption{Design of phase-change metasurface for dynamic radial metalens at distinct wavelengths. (a) Schematic diagram of the dynamic narrowband focusing in 2D space. (b) The spatial phase profile for the 2D focusing. (c) Transmission spectra of the phase-change nonlocal metasurface at different crystallization states. (d) The transverse far-field profiles on the focal plane and the longitudinal far-field profiles through the focal spot, for the converted LCP component at the corresponding resonant wavelength of a-Sb$_2$S$_3$, semi-Sb$_2$S$_3$, and c-Sb$_2$S$_3$ metasurfaces.}
	\label{Fig5}
\end{figure}

The transmission spectra of the designed metasurface for different crystallization states of Sb$_2$S$_3$ are depicted in Fig. 5(c). Compared with those of beam deflection metasurface and the cylindrical metalens above, the resonances remain intact even through the various geometry in these metadevices. This implies that the quasi-BIC resonance has a relatively stable resonance wavelength and peak amplitude for the various combinations of specific meta-atoms either in 1D or 2D. Such robustness would benefit the metadevice design where the spectral response can be defined by designing a single unit cell. The transverse and longitudinal far-field distributions with the focal spots on the quasi-BIC resonances in Fig. 5(d) show the efficient focusing ability at the corresponding resonant wavelength of different Sb$_2$S$_3$ metasurfaces. The relatively larger focusing intensity of c-Sb$_2$S$_3$ metasurface can be attributed to the increasing refractive index during the crystallization process. In addition, the focal spots off-resonance for c-Sb$_2$S$_3$ show the efficiency decreasing as the wavelengths gradually move away from the center of the resonance (shown in Figure S5). With several orders of magnitude smaller power at off-resonance wavelengths than background plane wave, the meatless can be considered functionally transparent off-resonance. Such spectral selectivity to the crystallization state of Sb$_2$S$_3$ leads to the tunability to either manipulate wavefront at a controllable wavelength or switch on and off the function at a selected wavelength.

In the proposed phase-change nonlocal metasurface, the flexibility to control both the spatial and spectral response can provide abundant degrees of freedom for implementing the narrowband wavefront manipulation in various applications associated with optical information processing. To demonstrate the versatility and high performance, a phase-only meta-hologram is finally designed. Using the same meta-atom library with the lenses above, the information is coded by the geometric phases of Sb$_2$S$_3$ meta-atoms that are exerted to the transmitted converted component on resonance. Here, the phase profile of the image is obtained by the Gerchberg–Saxton algorithm and then encoded by the 2D spatial distribution of the rotation angle of Sb$_2$S$_3$ meta-atoms. 

\begin{figure}[htbp]
	\centering
	\includegraphics[scale=0.4]{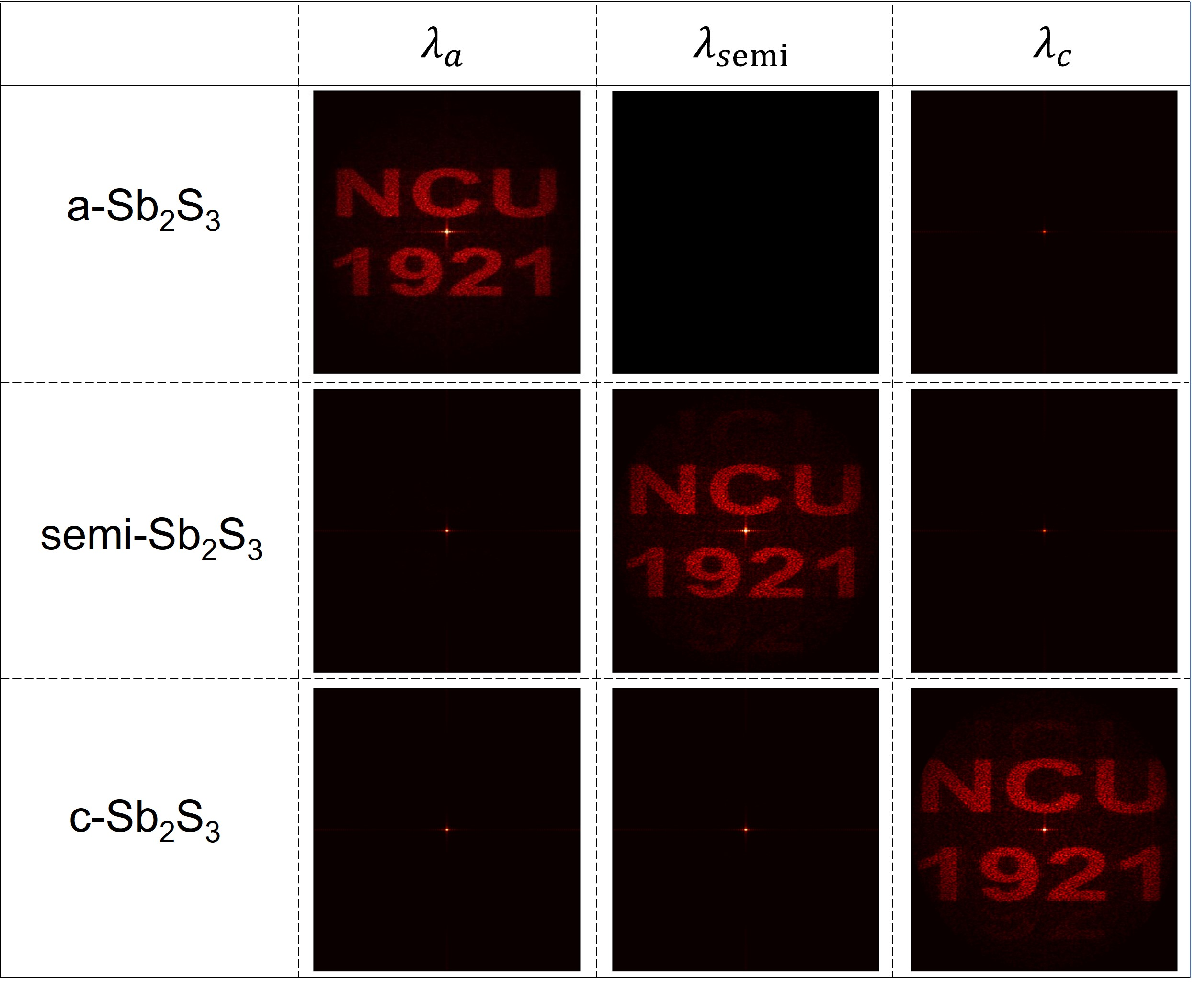}
	\caption{Demonstrations of dynamic spatial and spectral control in the phase-change nonlocal metasurfaces using a phase-only meta-hologram. Under RCP illumination, the diffraction patterns generated by the a-Sb$_2$S$_3$, semi-Sb$_2$S$_3$, and c-Sb$_2$S$_3$ metasurfaces of transmitted LCP component displays images exclusively at selected wavelengths and display misleading or useless information for the non-resonant regime.}
	\label{fig6}
\end{figure}

The simulated diffraction patterns shown in Fig. 6 are generated by the a-Sb$_2$S$_3$, semi-Sb$_2$S$_3$, and c-Sb$_2$S$_3$, respectively. For comparison, the holographic images at resonant and non-resonant wavelengths are presented. Before the phase transition of Sb$_2$S$_3$ meta-atoms, namely a-Sb$_2$S$_3$ metasurface, the image is only captured for the converted LCP light at the corresponding resonant wavelength around 1380 nm, while no image displays at the other two non-resonant wavelengths. Similar results can be observed for the semi-Sb$_2$S$_3$ and c-Sb$_2$S$_3$ metasurfaces. Such a mapping relationship in the space-wavelength domains for holographic imaging originated from the fact that the geometric phase profiles in the metasurface are fundamentally robust at different resonant wavelengths, with the correlated rotation angles of meta-atoms hardly affecting the spectral response, as shown in Fig. 2(f). These significant properties can be applied to optical information encryption where the information decryption requires the matching conditions between the selected wavelength and the appropriate crystallization state of Sb$_2$S$_3$ meta-atoms. More importantly, the proposed metasurface attempts to develop the optical metadevice capable of performing dynamic operations on incident light waves with high degrees of spectral control. The proposed metasurface allows the transmitted information only at selected narrowband wavelengths to be processed, avoiding the extra optical components such as polarizers to add the size and mass of system. It can also achieve an adaptive selection of the operating wavelength across a broad spectral bandwidth for information imaging, display, and sensing. This may open exciting opportunities in the quest to tailor light in spectral and spatial control using ultra-thin platforms for future AR/VR wearable devices.

\section{\label{sec4}Conclusions}

In conclusion, we have demonstrated phase-change nonlocal metasurfaces capable of actively shaping light's spatial and spectral properties in transmission mode. The functionality is enabled by incorporating the phase-change meta-atoms integrated with a nonlocal wavefront-shaping metasurface empowered by quasi-BICs. A systematic design approach is presented by combining the nonlocal design strategies of enhanced light-matter interaction of quasi-BICs and the wavefront-shaping principle encapsulated by the generalized Snell’s law in the phase-change metasurfaces. Leveraging spatially varying the polarization properties of quasi-BICs for designer geometric phase profiles, the wavefront-shaping functions for beam steering and space focusing are demonstrated for a narrow spectral regime in Sb$_2$S$_3$ metasurfaces. By switching the state of Sb$_2$S$_3$ between amorphous and crystalline, the wavefront manipulation can be realized at multiwavelength within a broad spectral range, and the spatial response at a specific wavelength can be switched on to off. The presented Sb$_2$S$_3$ metasurface platform can afford multistate, reversible, and nonvolatile tuning of the spatial and spectral response, offering sufficient design freedom to tailor the spatial response at tunable wavelength actively. Furthermore, the dynamic control over a single device's spatial and spectral response enables holographic imaging at selected wavelengths, which may open the door for next-generation compact narrowband imaging for applications such as information storage and encryption and AR/VR.

\begin{acknowledgments}	
	
This work was supported by the National Natural Science Foundation of China (Grants No. 12364045, 12304420, 12264028, and 12104105), the Natural Science Foundation of Jiangxi Province (Grants No. 20232BAB211025 and 20232BAB201040), the Chenguang Program of Shanghai Education Development Foundation and Shanghai Municipal Education Commission (Grant No. 21CGA55), the Shanghai Pujiang Program (Grant No. 22PJ1402900), the Guangdong Basic and Applied Basic Research Foundation (Grant No. 2023A1515011024), the Science and Technology Program of Guangzhou (Grant No. 202201011176), the China Scholarship Council (Grant No. 202008420045), and the Australian Research Council Discovery Project (Grant No. DP200101353). 

\end{acknowledgments}


\begin{thebibliography}{54}%
	\makeatletter
	\providecommand \@ifxundefined [1]{%
		\@ifx{#1\undefined}
	}%
	\providecommand \@ifnum [1]{%
		\ifnum #1\expandafter \@firstoftwo
		\else \expandafter \@secondoftwo
		\fi
	}%
	\providecommand \@ifx [1]{%
		\ifx #1\expandafter \@firstoftwo
		\else \expandafter \@secondoftwo
		\fi
	}%
	\providecommand \natexlab [1]{#1}%
	\providecommand \enquote  [1]{``#1''}%
	\providecommand \bibnamefont  [1]{#1}%
	\providecommand \bibfnamefont [1]{#1}%
	\providecommand \citenamefont [1]{#1}%
	\providecommand \href@noop [0]{\@secondoftwo}%
	\providecommand \href [0]{\begingroup \@sanitize@url \@href}%
	\providecommand \@href[1]{\@@startlink{#1}\@@href}%
	\providecommand \@@href[1]{\endgroup#1\@@endlink}%
	\providecommand \@sanitize@url [0]{\catcode `\\12\catcode `\$12\catcode
		`\&12\catcode `\#12\catcode `\^12\catcode `\_12\catcode `\%12\relax}%
	\providecommand \@@startlink[1]{}%
	\providecommand \@@endlink[0]{}%
	\providecommand \url  [0]{\begingroup\@sanitize@url \@url }%
	\providecommand \@url [1]{\endgroup\@href {#1}{\urlprefix }}%
	\providecommand \urlprefix  [0]{URL }%
	\providecommand \Eprint [0]{\href }%
	\providecommand \doibase [0]{https://doi.org/}%
	\providecommand \selectlanguage [0]{\@gobble}%
	\providecommand \bibinfo  [0]{\@secondoftwo}%
	\providecommand \bibfield  [0]{\@secondoftwo}%
	\providecommand \translation [1]{[#1]}%
	\providecommand \BibitemOpen [0]{}%
	\providecommand \bibitemStop [0]{}%
	\providecommand \bibitemNoStop [0]{.\EOS\space}%
	\providecommand \EOS [0]{\spacefactor3000\relax}%
	\providecommand \BibitemShut  [1]{\csname bibitem#1\endcsname}%
	\let\auto@bib@innerbib\@empty
	\bibitem [{\citenamefont {Kildishev}\ \emph {et~al.}(2013)\citenamefont
		{Kildishev}, \citenamefont {Boltasseva},\ and\ \citenamefont
		{Shalaev}}]{Kildishev2013}%
	\BibitemOpen
	\bibfield  {author} {\bibinfo {author} {\bibfnamefont {A.~V.}\ \bibnamefont
			{Kildishev}}, \bibinfo {author} {\bibfnamefont {A.}~\bibnamefont
			{Boltasseva}},\ and\ \bibinfo {author} {\bibfnamefont {V.~M.}\ \bibnamefont
			{Shalaev}},\ }\bibfield  {title} {\bibinfo {title} {Planar photonics with
			metasurfaces},\ }\href {https://doi.org/10.1126/science.1232009} {\bibfield
		{journal} {\bibinfo  {journal} {Science}\ }\textbf {\bibinfo {volume}
			{339}},\ \bibinfo {pages} {1232009} (\bibinfo {year} {2013})}\BibitemShut
	{NoStop}%
	\bibitem [{\citenamefont {Yu}\ and\ \citenamefont {Capasso}(2014)}]{Yu2014}%
	\BibitemOpen
	\bibfield  {author} {\bibinfo {author} {\bibfnamefont {N.}~\bibnamefont
			{Yu}}\ and\ \bibinfo {author} {\bibfnamefont {F.}~\bibnamefont {Capasso}},\
	}\bibfield  {title} {\bibinfo {title} {Flat optics with designer
			metasurfaces},\ }\href {https://doi.org/10.1038/nmat3839} {\bibfield
		{journal} {\bibinfo  {journal} {Nat. Mater.}\ }\textbf {\bibinfo {volume}
			{13}},\ \bibinfo {pages} {139} (\bibinfo {year} {2014})}\BibitemShut
	{NoStop}%
	\bibitem [{\citenamefont {Ding}\ \emph {et~al.}(2017)\citenamefont {Ding},
		\citenamefont {Pors},\ and\ \citenamefont {Bozhevolnyi}}]{Ding2017}%
	\BibitemOpen
	\bibfield  {author} {\bibinfo {author} {\bibfnamefont {F.}~\bibnamefont
			{Ding}}, \bibinfo {author} {\bibfnamefont {A.}~\bibnamefont {Pors}},\ and\
		\bibinfo {author} {\bibfnamefont {S.~I.}\ \bibnamefont {Bozhevolnyi}},\
	}\bibfield  {title} {\bibinfo {title} {Gradient metasurfaces: a review of
			fundamentals and applications},\ }\href
	{https://doi.org/10.1088/1361-6633/aa8732} {\bibfield  {journal} {\bibinfo
			{journal} {Rep. Prog. Phys.}\ }\textbf {\bibinfo {volume} {81}},\ \bibinfo
		{pages} {026401} (\bibinfo {year} {2017})}\BibitemShut {NoStop}%
	\bibitem [{\citenamefont {So}\ \emph {et~al.}(2023)\citenamefont {So},
		\citenamefont {Mun}, \citenamefont {Park},\ and\ \citenamefont
		{Rho}}]{So2023}%
	\BibitemOpen
	\bibfield  {author} {\bibinfo {author} {\bibfnamefont {S.}~\bibnamefont
			{So}}, \bibinfo {author} {\bibfnamefont {J.}~\bibnamefont {Mun}}, \bibinfo
		{author} {\bibfnamefont {J.}~\bibnamefont {Park}},\ and\ \bibinfo {author}
		{\bibfnamefont {J.}~\bibnamefont {Rho}},\ }\bibfield  {title} {\bibinfo
		{title} {Revisiting the design strategies for metasurfaces: Fundamental
			physics, optimization, and beyond},\ }\href
	{https://doi.org/10.1002/adma.202206399} {\bibfield  {journal} {\bibinfo
			{journal} {Adv. Mater.}\ ,\ \bibinfo {pages} {2206399}} (\bibinfo {year}
		{2023})}\BibitemShut {NoStop}%
	\bibitem [{\citenamefont {Arbabi}\ and\ \citenamefont
		{Faraon}(2022)}]{Arbabi2022}%
	\BibitemOpen
	\bibfield  {author} {\bibinfo {author} {\bibfnamefont {A.}~\bibnamefont
			{Arbabi}}\ and\ \bibinfo {author} {\bibfnamefont {A.}~\bibnamefont
			{Faraon}},\ }\bibfield  {title} {\bibinfo {title} {Advances in optical
			metalenses},\ }\href {https://doi.org/10.1038/s41566-022-01108-6} {\bibfield
		{journal} {\bibinfo  {journal} {Nat. Photonics}\ }\textbf {\bibinfo {volume}
			{17}},\ \bibinfo {pages} {16} (\bibinfo {year} {2022})}\BibitemShut {NoStop}%
	\bibitem [{\citenamefont {Yu}\ \emph {et~al.}(2011)\citenamefont {Yu},
		\citenamefont {Genevet}, \citenamefont {Kats}, \citenamefont {Aieta},
		\citenamefont {Tetienne}, \citenamefont {Capasso},\ and\ \citenamefont
		{Gaburro}}]{yu2011light}%
	\BibitemOpen
	\bibfield  {author} {\bibinfo {author} {\bibfnamefont {N.}~\bibnamefont
			{Yu}}, \bibinfo {author} {\bibfnamefont {P.}~\bibnamefont {Genevet}},
		\bibinfo {author} {\bibfnamefont {M.~A.}\ \bibnamefont {Kats}}, \bibinfo
		{author} {\bibfnamefont {F.}~\bibnamefont {Aieta}}, \bibinfo {author}
		{\bibfnamefont {J.-P.}\ \bibnamefont {Tetienne}}, \bibinfo {author}
		{\bibfnamefont {F.}~\bibnamefont {Capasso}},\ and\ \bibinfo {author}
		{\bibfnamefont {Z.}~\bibnamefont {Gaburro}},\ }\bibfield  {title} {\bibinfo
		{title} {Light propagation with phase discontinuities: generalized laws of
			reflection and refraction},\ }\href@noop {} {\bibfield  {journal} {\bibinfo
			{journal} {Science}\ }\textbf {\bibinfo {volume} {334}},\ \bibinfo {pages}
		{333} (\bibinfo {year} {2011})}\BibitemShut {NoStop}%
	\bibitem [{\citenamefont {Sun}\ \emph {et~al.}(2012)\citenamefont {Sun},
		\citenamefont {He}, \citenamefont {Xiao}, \citenamefont {Xu}, \citenamefont
		{Li},\ and\ \citenamefont {Zhou}}]{sun2012gradient}%
	\BibitemOpen
	\bibfield  {author} {\bibinfo {author} {\bibfnamefont {S.}~\bibnamefont
			{Sun}}, \bibinfo {author} {\bibfnamefont {Q.}~\bibnamefont {He}}, \bibinfo
		{author} {\bibfnamefont {S.}~\bibnamefont {Xiao}}, \bibinfo {author}
		{\bibfnamefont {Q.}~\bibnamefont {Xu}}, \bibinfo {author} {\bibfnamefont
			{X.}~\bibnamefont {Li}},\ and\ \bibinfo {author} {\bibfnamefont
			{L.}~\bibnamefont {Zhou}},\ }\bibfield  {title} {\bibinfo {title}
		{Gradient-index meta-surfaces as a bridge linking propagating waves and
			surface waves},\ }\href@noop {} {\bibfield  {journal} {\bibinfo  {journal}
			{Nat. Mater.}\ }\textbf {\bibinfo {volume} {11}},\ \bibinfo {pages} {426}
		(\bibinfo {year} {2012})}\BibitemShut {NoStop}%
	\bibitem [{\citenamefont {Chen}\ \emph {et~al.}(2018)\citenamefont {Chen},
		\citenamefont {Zhu}, \citenamefont {Sanjeev}, \citenamefont {Khorasaninejad},
		\citenamefont {Shi}, \citenamefont {Lee},\ and\ \citenamefont
		{Capasso}}]{Chen2018}%
	\BibitemOpen
	\bibfield  {author} {\bibinfo {author} {\bibfnamefont {W.~T.}\ \bibnamefont
			{Chen}}, \bibinfo {author} {\bibfnamefont {A.~Y.}\ \bibnamefont {Zhu}},
		\bibinfo {author} {\bibfnamefont {V.}~\bibnamefont {Sanjeev}}, \bibinfo
		{author} {\bibfnamefont {M.}~\bibnamefont {Khorasaninejad}}, \bibinfo
		{author} {\bibfnamefont {Z.}~\bibnamefont {Shi}}, \bibinfo {author}
		{\bibfnamefont {E.}~\bibnamefont {Lee}},\ and\ \bibinfo {author}
		{\bibfnamefont {F.}~\bibnamefont {Capasso}},\ }\bibfield  {title} {\bibinfo
		{title} {A broadband achromatic metalens for focusing and imaging in the
			visible},\ }\href {https://doi.org/10.1038/s41565-017-0034-6} {\bibfield
		{journal} {\bibinfo  {journal} {Nat. Nanotechnol.}\ }\textbf {\bibinfo
			{volume} {13}},\ \bibinfo {pages} {220} (\bibinfo {year} {2018})}\BibitemShut
	{NoStop}%
	\bibitem [{\citenamefont {Lin}\ \emph {et~al.}(2019)\citenamefont {Lin},
		\citenamefont {Su}, \citenamefont {Wang}, \citenamefont {Chen}, \citenamefont
		{Chung}, \citenamefont {Chen}, \citenamefont {Kuo}, \citenamefont {Chen},
		\citenamefont {Chen}, \citenamefont {Huang}, \citenamefont {Wang},
		\citenamefont {Chu}, \citenamefont {Wu}, \citenamefont {Li}, \citenamefont
		{Wang}, \citenamefont {Zhu},\ and\ \citenamefont {Tsai}}]{Lin2019}%
	\BibitemOpen
	\bibfield  {author} {\bibinfo {author} {\bibfnamefont {R.~J.}\ \bibnamefont
			{Lin}}, \bibinfo {author} {\bibfnamefont {V.-C.}\ \bibnamefont {Su}},
		\bibinfo {author} {\bibfnamefont {S.}~\bibnamefont {Wang}}, \bibinfo {author}
		{\bibfnamefont {M.~K.}\ \bibnamefont {Chen}}, \bibinfo {author}
		{\bibfnamefont {T.~L.}\ \bibnamefont {Chung}}, \bibinfo {author}
		{\bibfnamefont {Y.~H.}\ \bibnamefont {Chen}}, \bibinfo {author}
		{\bibfnamefont {H.~Y.}\ \bibnamefont {Kuo}}, \bibinfo {author} {\bibfnamefont
			{J.-W.}\ \bibnamefont {Chen}}, \bibinfo {author} {\bibfnamefont
			{J.}~\bibnamefont {Chen}}, \bibinfo {author} {\bibfnamefont {Y.-T.}\
			\bibnamefont {Huang}}, \bibinfo {author} {\bibfnamefont {J.-H.}\ \bibnamefont
			{Wang}}, \bibinfo {author} {\bibfnamefont {C.~H.}\ \bibnamefont {Chu}},
		\bibinfo {author} {\bibfnamefont {P.~C.}\ \bibnamefont {Wu}}, \bibinfo
		{author} {\bibfnamefont {T.}~\bibnamefont {Li}}, \bibinfo {author}
		{\bibfnamefont {Z.}~\bibnamefont {Wang}}, \bibinfo {author} {\bibfnamefont
			{S.}~\bibnamefont {Zhu}},\ and\ \bibinfo {author} {\bibfnamefont {D.~P.}\
			\bibnamefont {Tsai}},\ }\bibfield  {title} {\bibinfo {title} {Achromatic
			metalens array for full-colour light-field imaging},\ }\href
	{https://doi.org/10.1038/s41565-018-0347-0} {\bibfield  {journal} {\bibinfo
			{journal} {Nat. Nanotechnol.}\ }\textbf {\bibinfo {volume} {14}},\ \bibinfo
		{pages} {227} (\bibinfo {year} {2019})}\BibitemShut {NoStop}%
	\bibitem [{\citenamefont {Li}\ \emph {et~al.}(2020)\citenamefont {Li},
		\citenamefont {Chen}, \citenamefont {Guan}, \citenamefont {Tao},
		\citenamefont {Chang}, \citenamefont {Dai}, \citenamefont {Xiao},
		\citenamefont {Cui}, \citenamefont {Wang}, \citenamefont {Yu} \emph
		{et~al.}}]{li2020three}%
	\BibitemOpen
	\bibfield  {author} {\bibinfo {author} {\bibfnamefont {Z.}~\bibnamefont
			{Li}}, \bibinfo {author} {\bibfnamefont {C.}~\bibnamefont {Chen}}, \bibinfo
		{author} {\bibfnamefont {Z.}~\bibnamefont {Guan}}, \bibinfo {author}
		{\bibfnamefont {J.}~\bibnamefont {Tao}}, \bibinfo {author} {\bibfnamefont
			{S.}~\bibnamefont {Chang}}, \bibinfo {author} {\bibfnamefont
			{Q.}~\bibnamefont {Dai}}, \bibinfo {author} {\bibfnamefont {Y.}~\bibnamefont
			{Xiao}}, \bibinfo {author} {\bibfnamefont {Y.}~\bibnamefont {Cui}}, \bibinfo
		{author} {\bibfnamefont {Y.}~\bibnamefont {Wang}}, \bibinfo {author}
		{\bibfnamefont {S.}~\bibnamefont {Yu}}, \emph {et~al.},\ }\bibfield  {title}
	{\bibinfo {title} {Three-channel metasurfaces for simultaneous
			meta-holography and meta-nanoprinting: a single-cell design approach},\
	}\href@noop {} {\bibfield  {journal} {\bibinfo  {journal} {Laser Photonics
				Rev.}\ }\textbf {\bibinfo {volume} {14}},\ \bibinfo {pages} {2000032}
		(\bibinfo {year} {2020})}\BibitemShut {NoStop}%
	\bibitem [{\citenamefont {Georgi}\ \emph {et~al.}(2021)\citenamefont {Georgi},
		\citenamefont {Wei}, \citenamefont {Sain}, \citenamefont {Schlickriede},
		\citenamefont {Wang}, \citenamefont {Huang},\ and\ \citenamefont
		{Zentgraf}}]{Georgi2021}%
	\BibitemOpen
	\bibfield  {author} {\bibinfo {author} {\bibfnamefont {P.}~\bibnamefont
			{Georgi}}, \bibinfo {author} {\bibfnamefont {Q.}~\bibnamefont {Wei}},
		\bibinfo {author} {\bibfnamefont {B.}~\bibnamefont {Sain}}, \bibinfo {author}
		{\bibfnamefont {C.}~\bibnamefont {Schlickriede}}, \bibinfo {author}
		{\bibfnamefont {Y.}~\bibnamefont {Wang}}, \bibinfo {author} {\bibfnamefont
			{L.}~\bibnamefont {Huang}},\ and\ \bibinfo {author} {\bibfnamefont
			{T.}~\bibnamefont {Zentgraf}},\ }\bibfield  {title} {\bibinfo {title}
		{Optical secret sharing with cascaded metasurface holography},\ }\href
	{https://doi.org/10.1126/sciadv.abf9718} {\bibfield  {journal} {\bibinfo
			{journal} {Sci. Adv.}\ }\textbf {\bibinfo {volume} {7}},\ \bibinfo {pages}
		{eabf9718} (\bibinfo {year} {2021})}\BibitemShut {NoStop}%
	\bibitem [{\citenamefont {Guo}\ \emph {et~al.}(2021)\citenamefont {Guo},
		\citenamefont {Zhang}, \citenamefont {Pu}, \citenamefont {He}, \citenamefont
		{Jin}, \citenamefont {Xu}, \citenamefont {Zhang}, \citenamefont {Gao},\ and\
		\citenamefont {Luo}}]{Guo2021}%
	\BibitemOpen
	\bibfield  {author} {\bibinfo {author} {\bibfnamefont {Y.}~\bibnamefont
			{Guo}}, \bibinfo {author} {\bibfnamefont {S.}~\bibnamefont {Zhang}}, \bibinfo
		{author} {\bibfnamefont {M.}~\bibnamefont {Pu}}, \bibinfo {author}
		{\bibfnamefont {Q.}~\bibnamefont {He}}, \bibinfo {author} {\bibfnamefont
			{J.}~\bibnamefont {Jin}}, \bibinfo {author} {\bibfnamefont {M.}~\bibnamefont
			{Xu}}, \bibinfo {author} {\bibfnamefont {Y.}~\bibnamefont {Zhang}}, \bibinfo
		{author} {\bibfnamefont {P.}~\bibnamefont {Gao}},\ and\ \bibinfo {author}
		{\bibfnamefont {X.}~\bibnamefont {Luo}},\ }\bibfield  {title} {\bibinfo
		{title} {Spin-decoupled metasurface for simultaneous detection of spin and
			orbital angular momenta via momentum transformation},\ }\href
	{https://doi.org/10.1038/s41377-021-00497-7} {\bibfield  {journal} {\bibinfo
			{journal} {Light Sci. Appl.}\ }\textbf {\bibinfo {volume} {10}},\ \bibinfo
		{pages} {63} (\bibinfo {year} {2021})}\BibitemShut {NoStop}%
	\bibitem [{\citenamefont {Luo}\ \emph {et~al.}(2022)\citenamefont {Luo},
		\citenamefont {Hu}, \citenamefont {Ou}, \citenamefont {Li}, \citenamefont
		{Lai}, \citenamefont {Liu}, \citenamefont {Cheng}, \citenamefont {Pan},\ and\
		\citenamefont {Duan}}]{Luo2022}%
	\BibitemOpen
	\bibfield  {author} {\bibinfo {author} {\bibfnamefont {X.}~\bibnamefont
			{Luo}}, \bibinfo {author} {\bibfnamefont {Y.}~\bibnamefont {Hu}}, \bibinfo
		{author} {\bibfnamefont {X.}~\bibnamefont {Ou}}, \bibinfo {author}
		{\bibfnamefont {X.}~\bibnamefont {Li}}, \bibinfo {author} {\bibfnamefont
			{J.}~\bibnamefont {Lai}}, \bibinfo {author} {\bibfnamefont {N.}~\bibnamefont
			{Liu}}, \bibinfo {author} {\bibfnamefont {X.}~\bibnamefont {Cheng}}, \bibinfo
		{author} {\bibfnamefont {A.}~\bibnamefont {Pan}},\ and\ \bibinfo {author}
		{\bibfnamefont {H.}~\bibnamefont {Duan}},\ }\bibfield  {title} {\bibinfo
		{title} {Metasurface-enabled on-chip multiplexed diffractive neural networks
			in the visible},\ }\href {https://doi.org/10.1038/s41377-022-00844-2}
	{\bibfield  {journal} {\bibinfo  {journal} {Light Sci. Appl.}\ }\textbf
		{\bibinfo {volume} {11}},\ \bibinfo {pages} {158} (\bibinfo {year}
		{2022})}\BibitemShut {NoStop}%
	\bibitem [{\citenamefont {Ren}\ \emph {et~al.}(2020)\citenamefont {Ren},
		\citenamefont {Fang}, \citenamefont {Jang}, \citenamefont {B{\"u}rger},
		\citenamefont {Rho},\ and\ \citenamefont {Maier}}]{ren2020complex}%
	\BibitemOpen
	\bibfield  {author} {\bibinfo {author} {\bibfnamefont {H.}~\bibnamefont
			{Ren}}, \bibinfo {author} {\bibfnamefont {X.}~\bibnamefont {Fang}}, \bibinfo
		{author} {\bibfnamefont {J.}~\bibnamefont {Jang}}, \bibinfo {author}
		{\bibfnamefont {J.}~\bibnamefont {B{\"u}rger}}, \bibinfo {author}
		{\bibfnamefont {J.}~\bibnamefont {Rho}},\ and\ \bibinfo {author}
		{\bibfnamefont {S.~A.}\ \bibnamefont {Maier}},\ }\bibfield  {title} {\bibinfo
		{title} {Complex-amplitude metasurface-based orbital angular momentum
			holography in momentum space},\ }\href@noop {} {\bibfield  {journal}
		{\bibinfo  {journal} {Nat. Nanotechnol.}\ }\textbf {\bibinfo {volume} {15}},\
		\bibinfo {pages} {948} (\bibinfo {year} {2020})}\BibitemShut {NoStop}%
	\bibitem [{\citenamefont {Li}\ and\ \citenamefont {Hsu}(2022)}]{Li2022a}%
	\BibitemOpen
	\bibfield  {author} {\bibinfo {author} {\bibfnamefont {S.}~\bibnamefont
			{Li}}\ and\ \bibinfo {author} {\bibfnamefont {C.~W.}\ \bibnamefont {Hsu}},\
	}\bibfield  {title} {\bibinfo {title} {Thickness bound for nonlocal
			wide-field-of-view metalenses},\ }\href
	{https://doi.org/10.1038/s41377-022-01038-6} {\bibfield  {journal} {\bibinfo
			{journal} {Light Sci. Appl.}\ }\textbf {\bibinfo {volume} {11}},\ \bibinfo
		{pages} {338} (\bibinfo {year} {2022})}\BibitemShut {NoStop}%
	\bibitem [{\citenamefont {Li}\ and\ \citenamefont {Hsu}(2023)}]{Li2023}%
	\BibitemOpen
	\bibfield  {author} {\bibinfo {author} {\bibfnamefont {S.}~\bibnamefont
			{Li}}\ and\ \bibinfo {author} {\bibfnamefont {C.~W.}\ \bibnamefont {Hsu}},\
	}\bibfield  {title} {\bibinfo {title} {Transmission efficiency limit for
			nonlocal metalenses},\ }\href {https://doi.org/10.1002/lpor.202300201}
	{\bibfield  {journal} {\bibinfo  {journal} {Laser Photonics Rev.}\ ,\
			\bibinfo {pages} {2300201}} (\bibinfo {year} {2023})}\BibitemShut {NoStop}%
	\bibitem [{\citenamefont {Overvig}\ \emph
		{et~al.}(2020{\natexlab{a}})\citenamefont {Overvig}, \citenamefont {Malek},
		\citenamefont {Carter}, \citenamefont {Shrestha},\ and\ \citenamefont
		{Yu}}]{Overvig2020}%
	\BibitemOpen
	\bibfield  {author} {\bibinfo {author} {\bibfnamefont {A.~C.}\ \bibnamefont
			{Overvig}}, \bibinfo {author} {\bibfnamefont {S.~C.}\ \bibnamefont {Malek}},
		\bibinfo {author} {\bibfnamefont {M.~J.}\ \bibnamefont {Carter}}, \bibinfo
		{author} {\bibfnamefont {S.}~\bibnamefont {Shrestha}},\ and\ \bibinfo
		{author} {\bibfnamefont {N.}~\bibnamefont {Yu}},\ }\bibfield  {title}
	{\bibinfo {title} {Selection rules for quasibound states in the continuum},\
	}\href {https://doi.org/10.1103/physrevb.102.035434} {\bibfield  {journal}
		{\bibinfo  {journal} {Phys. Rev. B}\ }\textbf {\bibinfo {volume} {102}},\
		\bibinfo {pages} {035434} (\bibinfo {year} {2020}{\natexlab{a}})}\BibitemShut
	{NoStop}%
	\bibitem [{\citenamefont {Sang}\ \emph {et~al.}(2022)\citenamefont {Sang},
		\citenamefont {Xu}, \citenamefont {An},\ and\ \citenamefont {Fu}}]{Sang2022}%
	\BibitemOpen
	\bibfield  {author} {\bibinfo {author} {\bibfnamefont {D.}~\bibnamefont
			{Sang}}, \bibinfo {author} {\bibfnamefont {M.}~\bibnamefont {Xu}}, \bibinfo
		{author} {\bibfnamefont {Q.}~\bibnamefont {An}},\ and\ \bibinfo {author}
		{\bibfnamefont {Y.}~\bibnamefont {Fu}},\ }\bibfield  {title} {\bibinfo
		{title} {Broadband transparent and high-q resonant polarization meta-grating
			enabled by a non-local geometric-phase metasurface},\ }\href
	{https://doi.org/10.1364/oe.462248} {\bibfield  {journal} {\bibinfo
			{journal} {Opt. Express}\ }\textbf {\bibinfo {volume} {30}},\ \bibinfo
		{pages} {26664} (\bibinfo {year} {2022})}\BibitemShut {NoStop}%
	\bibitem [{\citenamefont {Overvig}\ \emph {et~al.}(2023)\citenamefont
		{Overvig}, \citenamefont {Kasahara}, \citenamefont {Xu},\ and\ \citenamefont
		{Al{\`u}}}]{overvig2023demonstration}%
	\BibitemOpen
	\bibfield  {author} {\bibinfo {author} {\bibfnamefont {A.}~\bibnamefont
			{Overvig}}, \bibinfo {author} {\bibfnamefont {Y.}~\bibnamefont {Kasahara}},
		\bibinfo {author} {\bibfnamefont {G.}~\bibnamefont {Xu}},\ and\ \bibinfo
		{author} {\bibfnamefont {A.}~\bibnamefont {Al{\`u}}},\ }\bibfield  {title}
	{\bibinfo {title} {Demonstration of a polarization-agnostic geometric phase
			in nonlocal metasurfaces},\ }\href@noop {} {\bibfield  {journal} {\bibinfo
			{journal} {arXiv preprint arXiv:2302.13215}\ } (\bibinfo {year}
		{2023})}\BibitemShut {NoStop}%
	\bibitem [{\citenamefont {Huang}\ \emph
		{et~al.}(2023{\natexlab{a}})\citenamefont {Huang}, \citenamefont {Overvig},
		\citenamefont {Xu}, \citenamefont {Malek}, \citenamefont {Tsai},
		\citenamefont {Al{\`{u}}},\ and\ \citenamefont {Yu}}]{Huang2023}%
	\BibitemOpen
	\bibfield  {author} {\bibinfo {author} {\bibfnamefont {H.}~\bibnamefont
			{Huang}}, \bibinfo {author} {\bibfnamefont {A.~C.}\ \bibnamefont {Overvig}},
		\bibinfo {author} {\bibfnamefont {Y.}~\bibnamefont {Xu}}, \bibinfo {author}
		{\bibfnamefont {S.~C.}\ \bibnamefont {Malek}}, \bibinfo {author}
		{\bibfnamefont {C.-C.}\ \bibnamefont {Tsai}}, \bibinfo {author}
		{\bibfnamefont {A.}~\bibnamefont {Al{\`{u}}}},\ and\ \bibinfo {author}
		{\bibfnamefont {N.}~\bibnamefont {Yu}},\ }\bibfield  {title} {\bibinfo
		{title} {Leaky-wave metasurfaces for integrated photonics},\ }\href
	{https://doi.org/10.1038/s41565-023-01360-z} {\bibfield  {journal} {\bibinfo
			{journal} {Nat. Nanotechnol.}\ }\textbf {\bibinfo {volume} {18}},\ \bibinfo
		{pages} {580} (\bibinfo {year} {2023}{\natexlab{a}})}\BibitemShut {NoStop}%
	\bibitem [{\citenamefont {Xu}\ \emph {et~al.}(2023)\citenamefont {Xu},
		\citenamefont {Overvig}, \citenamefont {Kasahara}, \citenamefont {Martini},
		\citenamefont {Maci},\ and\ \citenamefont {Al{\`{u}}}}]{Xu2023}%
	\BibitemOpen
	\bibfield  {author} {\bibinfo {author} {\bibfnamefont {G.}~\bibnamefont
			{Xu}}, \bibinfo {author} {\bibfnamefont {A.}~\bibnamefont {Overvig}},
		\bibinfo {author} {\bibfnamefont {Y.}~\bibnamefont {Kasahara}}, \bibinfo
		{author} {\bibfnamefont {E.}~\bibnamefont {Martini}}, \bibinfo {author}
		{\bibfnamefont {S.}~\bibnamefont {Maci}},\ and\ \bibinfo {author}
		{\bibfnamefont {A.}~\bibnamefont {Al{\`{u}}}},\ }\bibfield  {title} {\bibinfo
		{title} {Arbitrary aperture synthesis with nonlocal leaky-wave metasurface
			antennas},\ }\href {https://doi.org/10.1038/s41467-023-39818-2} {\bibfield
		{journal} {\bibinfo  {journal} {Nat. Commun.}\ }\textbf {\bibinfo {volume}
			{14}},\ \bibinfo {pages} {4380} (\bibinfo {year} {2023})}\BibitemShut
	{NoStop}%
	\bibitem [{\citenamefont {Overvig}\ and\ \citenamefont
		{Al{\`{u}}}(2021)}]{Overvig2021}%
	\BibitemOpen
	\bibfield  {author} {\bibinfo {author} {\bibfnamefont {A.}~\bibnamefont
			{Overvig}}\ and\ \bibinfo {author} {\bibfnamefont {A.}~\bibnamefont
			{Al{\`{u}}}},\ }\bibfield  {title} {\bibinfo {title} {Wavefront-selective
			fano resonant metasurfaces},\ }\href
	{https://doi.org/10.1117/1.ap.3.2.026002} {\bibfield  {journal} {\bibinfo
			{journal} {Adv. Photonics}\ }\textbf {\bibinfo {volume} {3}},\ \bibinfo
		{pages} {026002} (\bibinfo {year} {2021})}\BibitemShut {NoStop}%
	\bibitem [{\citenamefont {Overvig}\ and\ \citenamefont
		{Al{\`{u}}}(2022)}]{Overvig2022}%
	\BibitemOpen
	\bibfield  {author} {\bibinfo {author} {\bibfnamefont {A.}~\bibnamefont
			{Overvig}}\ and\ \bibinfo {author} {\bibfnamefont {A.}~\bibnamefont
			{Al{\`{u}}}},\ }\bibfield  {title} {\bibinfo {title} {Diffractive nonlocal
			metasurfaces},\ }\href {https://doi.org/10.1002/lpor.202100633} {\bibfield
		{journal} {\bibinfo  {journal} {Laser Photonics Rev.}\ }\textbf {\bibinfo
			{volume} {16}},\ \bibinfo {pages} {2100633} (\bibinfo {year}
		{2022})}\BibitemShut {NoStop}%
	\bibitem [{\citenamefont {Chen}\ \emph {et~al.}(2022)\citenamefont {Chen},
		\citenamefont {Li}, \citenamefont {Bi}, \citenamefont {Wang}, \citenamefont
		{Zhu},\ and\ \citenamefont {Wang}}]{Chen2022}%
	\BibitemOpen
	\bibfield  {author} {\bibinfo {author} {\bibfnamefont {R.}~\bibnamefont
			{Chen}}, \bibinfo {author} {\bibfnamefont {T.}~\bibnamefont {Li}}, \bibinfo
		{author} {\bibfnamefont {Q.}~\bibnamefont {Bi}}, \bibinfo {author}
		{\bibfnamefont {S.}~\bibnamefont {Wang}}, \bibinfo {author} {\bibfnamefont
			{S.}~\bibnamefont {Zhu}},\ and\ \bibinfo {author} {\bibfnamefont
			{Z.}~\bibnamefont {Wang}},\ }\bibfield  {title} {\bibinfo {title}
		{Quasi-bound states in the continuum-based switchable light-field
			manipulator},\ }\href {https://doi.org/10.1364/ome.454022} {\bibfield
		{journal} {\bibinfo  {journal} {Opt. Mater. Express}\ }\textbf {\bibinfo
			{volume} {12}},\ \bibinfo {pages} {1232} (\bibinfo {year}
		{2022})}\BibitemShut {NoStop}%
	\bibitem [{\citenamefont {Malek}\ \emph {et~al.}(2022)\citenamefont {Malek},
		\citenamefont {Overvig}, \citenamefont {Al{\`{u}}},\ and\ \citenamefont
		{Yu}}]{Malek2022}%
	\BibitemOpen
	\bibfield  {author} {\bibinfo {author} {\bibfnamefont {S.~C.}\ \bibnamefont
			{Malek}}, \bibinfo {author} {\bibfnamefont {A.~C.}\ \bibnamefont {Overvig}},
		\bibinfo {author} {\bibfnamefont {A.}~\bibnamefont {Al{\`{u}}}},\ and\
		\bibinfo {author} {\bibfnamefont {N.}~\bibnamefont {Yu}},\ }\bibfield
	{title} {\bibinfo {title} {Multifunctional resonant wavefront-shaping
			meta-optics based on multilayer and multi-perturbation nonlocal
			metasurfaces},\ }\href {https://doi.org/10.1038/s41377-022-00905-6}
	{\bibfield  {journal} {\bibinfo  {journal} {Light Sci. Appl.}\ }\textbf
		{\bibinfo {volume} {11}},\ \bibinfo {pages} {246} (\bibinfo {year}
		{2022})}\BibitemShut {NoStop}%
	\bibitem [{\citenamefont {Overvig}\ \emph
		{et~al.}(2020{\natexlab{b}})\citenamefont {Overvig}, \citenamefont {Malek},\
		and\ \citenamefont {Yu}}]{Overvig2020a}%
	\BibitemOpen
	\bibfield  {author} {\bibinfo {author} {\bibfnamefont {A.~C.}\ \bibnamefont
			{Overvig}}, \bibinfo {author} {\bibfnamefont {S.~C.}\ \bibnamefont {Malek}},\
		and\ \bibinfo {author} {\bibfnamefont {N.}~\bibnamefont {Yu}},\ }\bibfield
	{title} {\bibinfo {title} {Multifunctional nonlocal metasurfaces},\ }\href
	{https://doi.org/10.1103/physrevlett.125.017402} {\bibfield  {journal}
		{\bibinfo  {journal} {Phys. Rev. Lett.}\ }\textbf {\bibinfo {volume} {125}},\
		\bibinfo {pages} {017402} (\bibinfo {year} {2020}{\natexlab{b}})}\BibitemShut
	{NoStop}%
	\bibitem [{\citenamefont {Zhou}\ \emph {et~al.}(2023)\citenamefont {Zhou},
		\citenamefont {Guo}, \citenamefont {Overvig},\ and\ \citenamefont
		{Al{\`{u}}}}]{Zhou2023}%
	\BibitemOpen
	\bibfield  {author} {\bibinfo {author} {\bibfnamefont {Y.}~\bibnamefont
			{Zhou}}, \bibinfo {author} {\bibfnamefont {S.}~\bibnamefont {Guo}}, \bibinfo
		{author} {\bibfnamefont {A.~C.}\ \bibnamefont {Overvig}},\ and\ \bibinfo
		{author} {\bibfnamefont {A.}~\bibnamefont {Al{\`{u}}}},\ }\bibfield  {title}
	{\bibinfo {title} {Multiresonant nonlocal metasurfaces},\ }\href
	{https://doi.org/10.1021/acs.nanolett.3c00772} {\bibfield  {journal}
		{\bibinfo  {journal} {Nano Lett.}\ }\textbf {\bibinfo {volume} {23}},\
		\bibinfo {pages} {6768} (\bibinfo {year} {2023})}\BibitemShut {NoStop}%
	\bibitem [{\citenamefont {Ding}\ \emph {et~al.}(2019)\citenamefont {Ding},
		\citenamefont {Yang},\ and\ \citenamefont {Bozhevolnyi}}]{Ding2019}%
	\BibitemOpen
	\bibfield  {author} {\bibinfo {author} {\bibfnamefont {F.}~\bibnamefont
			{Ding}}, \bibinfo {author} {\bibfnamefont {Y.}~\bibnamefont {Yang}},\ and\
		\bibinfo {author} {\bibfnamefont {S.~I.}\ \bibnamefont {Bozhevolnyi}},\
	}\bibfield  {title} {\bibinfo {title} {Dynamic metasurfaces using
			phase-change chalcogenides},\ }\href {https://doi.org/10.1002/adom.201801709}
	{\bibfield  {journal} {\bibinfo  {journal} {Adv. Opt. Mater.}\ }\textbf
		{\bibinfo {volume} {7}},\ \bibinfo {pages} {1801709} (\bibinfo {year}
		{2019})}\BibitemShut {NoStop}%
	\bibitem [{\citenamefont {Xiao}\ \emph {et~al.}(2020)\citenamefont {Xiao},
		\citenamefont {Wang}, \citenamefont {Liu}, \citenamefont {Zhou},
		\citenamefont {Jiang},\ and\ \citenamefont {Zhang}}]{Xiao2020}%
	\BibitemOpen
	\bibfield  {author} {\bibinfo {author} {\bibfnamefont {S.}~\bibnamefont
			{Xiao}}, \bibinfo {author} {\bibfnamefont {T.}~\bibnamefont {Wang}}, \bibinfo
		{author} {\bibfnamefont {T.}~\bibnamefont {Liu}}, \bibinfo {author}
		{\bibfnamefont {C.}~\bibnamefont {Zhou}}, \bibinfo {author} {\bibfnamefont
			{X.}~\bibnamefont {Jiang}},\ and\ \bibinfo {author} {\bibfnamefont
			{J.}~\bibnamefont {Zhang}},\ }\bibfield  {title} {\bibinfo {title} {Active
			metamaterials and metadevices: a review},\ }\href
	{https://doi.org/10.1088/1361-6463/abaced} {\bibfield  {journal} {\bibinfo
			{journal} {J. Phys. D: Appl. Phys.}\ }\textbf {\bibinfo {volume} {53}},\
		\bibinfo {pages} {503002} (\bibinfo {year} {2020})}\BibitemShut {NoStop}%
	\bibitem [{\citenamefont {Kim}\ \emph {et~al.}(2022)\citenamefont {Kim},
		\citenamefont {Seong}, \citenamefont {Yang}, \citenamefont {Moon},
		\citenamefont {Badloe},\ and\ \citenamefont {Rho}}]{Kim2022}%
	\BibitemOpen
	\bibfield  {author} {\bibinfo {author} {\bibfnamefont {J.}~\bibnamefont
			{Kim}}, \bibinfo {author} {\bibfnamefont {J.}~\bibnamefont {Seong}}, \bibinfo
		{author} {\bibfnamefont {Y.}~\bibnamefont {Yang}}, \bibinfo {author}
		{\bibfnamefont {S.-W.}\ \bibnamefont {Moon}}, \bibinfo {author}
		{\bibfnamefont {T.}~\bibnamefont {Badloe}},\ and\ \bibinfo {author}
		{\bibfnamefont {J.}~\bibnamefont {Rho}},\ }\bibfield  {title} {\bibinfo
		{title} {Tunable metasurfaces towards versatile metalenses and metaholograms:
			a review},\ }\href {https://doi.org/10.1117/1.ap.4.2.024001} {\bibfield
		{journal} {\bibinfo  {journal} {Adv. Photonics}\ }\textbf {\bibinfo {volume}
			{4}},\ \bibinfo {pages} {024001} (\bibinfo {year} {2022})}\BibitemShut
	{NoStop}%
	\bibitem [{\citenamefont {Abdelraouf}\ \emph {et~al.}(2022)\citenamefont
		{Abdelraouf}, \citenamefont {Wang}, \citenamefont {Liu}, \citenamefont
		{Dong}, \citenamefont {Wang}, \citenamefont {Ye}, \citenamefont {Wang},
		\citenamefont {Wang},\ and\ \citenamefont {Liu}}]{Abdelraouf2022}%
	\BibitemOpen
	\bibfield  {author} {\bibinfo {author} {\bibfnamefont {O.~A.~M.}\
			\bibnamefont {Abdelraouf}}, \bibinfo {author} {\bibfnamefont
			{Z.}~\bibnamefont {Wang}}, \bibinfo {author} {\bibfnamefont {H.}~\bibnamefont
			{Liu}}, \bibinfo {author} {\bibfnamefont {Z.}~\bibnamefont {Dong}}, \bibinfo
		{author} {\bibfnamefont {Q.}~\bibnamefont {Wang}}, \bibinfo {author}
		{\bibfnamefont {M.}~\bibnamefont {Ye}}, \bibinfo {author} {\bibfnamefont
			{X.~R.}\ \bibnamefont {Wang}}, \bibinfo {author} {\bibfnamefont {Q.~J.}\
			\bibnamefont {Wang}},\ and\ \bibinfo {author} {\bibfnamefont
			{H.}~\bibnamefont {Liu}},\ }\bibfield  {title} {\bibinfo {title} {Recent
			advances in tunable metasurfaces: Materials, design, and applications},\
	}\href {https://doi.org/10.1021/acsnano.2c04628} {\bibfield  {journal}
		{\bibinfo  {journal} {{ACS} Nano}\ }\textbf {\bibinfo {volume} {16}},\
		\bibinfo {pages} {13339} (\bibinfo {year} {2022})}\BibitemShut {NoStop}%
	\bibitem [{\citenamefont {Malek}\ \emph {et~al.}(2023)\citenamefont {Malek},
		\citenamefont {Tsai},\ and\ \citenamefont {Yu}}]{malek2023thermally}%
	\BibitemOpen
	\bibfield  {author} {\bibinfo {author} {\bibfnamefont {S.~C.}\ \bibnamefont
			{Malek}}, \bibinfo {author} {\bibfnamefont {C.-C.}\ \bibnamefont {Tsai}},\
		and\ \bibinfo {author} {\bibfnamefont {N.}~\bibnamefont {Yu}},\ }\bibfield
	{title} {\bibinfo {title} {Thermally-switchable metalenses based on
			quasi-bound states in the continuum},\ }\href@noop {} {\bibfield  {journal}
		{\bibinfo  {journal} {arXiv preprint arXiv:2306.13644}\ } (\bibinfo {year}
		{2023})}\BibitemShut {NoStop}%
	\bibitem [{\citenamefont {Malek}\ \emph {et~al.}(2020)\citenamefont {Malek},
		\citenamefont {Overvig}, \citenamefont {Shrestha},\ and\ \citenamefont
		{Yu}}]{Malek2020}%
	\BibitemOpen
	\bibfield  {author} {\bibinfo {author} {\bibfnamefont {S.~C.}\ \bibnamefont
			{Malek}}, \bibinfo {author} {\bibfnamefont {A.~C.}\ \bibnamefont {Overvig}},
		\bibinfo {author} {\bibfnamefont {S.}~\bibnamefont {Shrestha}},\ and\
		\bibinfo {author} {\bibfnamefont {N.}~\bibnamefont {Yu}},\ }\bibfield
	{title} {\bibinfo {title} {Active nonlocal metasurfaces},\ }\href
	{https://doi.org/10.1515/nanoph-2020-0375} {\bibfield  {journal} {\bibinfo
			{journal} {Nanophotonics}\ }\textbf {\bibinfo {volume} {10}},\ \bibinfo
		{pages} {655} (\bibinfo {year} {2020})}\BibitemShut {NoStop}%
	\bibitem [{\citenamefont {Hsu}\ \emph {et~al.}(2016)\citenamefont {Hsu},
		\citenamefont {Zhen}, \citenamefont {Stone}, \citenamefont {Joannopoulos},\
		and\ \citenamefont {Solja{\v{c}}i{\'{c}}}}]{Hsu2016}%
	\BibitemOpen
	\bibfield  {author} {\bibinfo {author} {\bibfnamefont {C.~W.}\ \bibnamefont
			{Hsu}}, \bibinfo {author} {\bibfnamefont {B.}~\bibnamefont {Zhen}}, \bibinfo
		{author} {\bibfnamefont {A.~D.}\ \bibnamefont {Stone}}, \bibinfo {author}
		{\bibfnamefont {J.~D.}\ \bibnamefont {Joannopoulos}},\ and\ \bibinfo {author}
		{\bibfnamefont {M.}~\bibnamefont {Solja{\v{c}}i{\'{c}}}},\ }\bibfield
	{title} {\bibinfo {title} {Bound states in the continuum},\ }\href
	{https://doi.org/10.1038/natrevmats.2016.48} {\bibfield  {journal} {\bibinfo
			{journal} {Nat. Rev. Mater.}\ }\textbf {\bibinfo {volume} {1}},\ \bibinfo
		{pages} {16048} (\bibinfo {year} {2016})}\BibitemShut {NoStop}%
	\bibitem [{\citenamefont {Sadreev}(2021)}]{Sadreev2021}%
	\BibitemOpen
	\bibfield  {author} {\bibinfo {author} {\bibfnamefont {A.~F.}\ \bibnamefont
			{Sadreev}},\ }\bibfield  {title} {\bibinfo {title} {Interference traps waves
			in an open system: bound states in the continuum},\ }\href
	{https://doi.org/10.1088/1361-6633/abefb9} {\bibfield  {journal} {\bibinfo
			{journal} {Rep. Prog. Phys.}\ }\textbf {\bibinfo {volume} {84}},\ \bibinfo
		{pages} {055901} (\bibinfo {year} {2021})}\BibitemShut {NoStop}%
	\bibitem [{\citenamefont {Huang}\ \emph
		{et~al.}(2023{\natexlab{b}})\citenamefont {Huang}, \citenamefont {Xu},
		\citenamefont {Powell}, \citenamefont {Padilla},\ and\ \citenamefont
		{Miroshnichenko}}]{Huang2023a}%
	\BibitemOpen
	\bibfield  {author} {\bibinfo {author} {\bibfnamefont {L.}~\bibnamefont
			{Huang}}, \bibinfo {author} {\bibfnamefont {L.}~\bibnamefont {Xu}}, \bibinfo
		{author} {\bibfnamefont {D.~A.}\ \bibnamefont {Powell}}, \bibinfo {author}
		{\bibfnamefont {W.~J.}\ \bibnamefont {Padilla}},\ and\ \bibinfo {author}
		{\bibfnamefont {A.~E.}\ \bibnamefont {Miroshnichenko}},\ }\bibfield  {title}
	{\bibinfo {title} {Resonant leaky modes in all-dielectric metasystems:
			Fundamentals and applications},\ }\href
	{https://doi.org/10.1016/j.physrep.2023.01.001} {\bibfield  {journal}
		{\bibinfo  {journal} {Phys. Rep.}\ }\textbf {\bibinfo {volume} {1008}},\
		\bibinfo {pages} {1} (\bibinfo {year} {2023}{\natexlab{b}})}\BibitemShut
	{NoStop}%
	\bibitem [{\citenamefont {Koshelev}\ \emph {et~al.}(2018)\citenamefont
		{Koshelev}, \citenamefont {Lepeshov}, \citenamefont {Liu}, \citenamefont
		{Bogdanov},\ and\ \citenamefont {Kivshar}}]{Koshelev2018}%
	\BibitemOpen
	\bibfield  {author} {\bibinfo {author} {\bibfnamefont {K.}~\bibnamefont
			{Koshelev}}, \bibinfo {author} {\bibfnamefont {S.}~\bibnamefont {Lepeshov}},
		\bibinfo {author} {\bibfnamefont {M.}~\bibnamefont {Liu}}, \bibinfo {author}
		{\bibfnamefont {A.}~\bibnamefont {Bogdanov}},\ and\ \bibinfo {author}
		{\bibfnamefont {Y.}~\bibnamefont {Kivshar}},\ }\bibfield  {title} {\bibinfo
		{title} {Asymmetric metasurfaces with high-q resonances governed by bound
			states in the continuum},\ }\href
	{https://doi.org/10.1103/physrevlett.121.193903} {\bibfield  {journal}
		{\bibinfo  {journal} {Phys. Rev. Lett.}\ }\textbf {\bibinfo {volume} {121}},\
		\bibinfo {pages} {193903} (\bibinfo {year} {2018})}\BibitemShut {NoStop}%
	\bibitem [{\citenamefont {Li}\ \emph {et~al.}(2019)\citenamefont {Li},
		\citenamefont {Zhou}, \citenamefont {Liu},\ and\ \citenamefont
		{Xiao}}]{Li2019}%
	\BibitemOpen
	\bibfield  {author} {\bibinfo {author} {\bibfnamefont {S.}~\bibnamefont
			{Li}}, \bibinfo {author} {\bibfnamefont {C.}~\bibnamefont {Zhou}}, \bibinfo
		{author} {\bibfnamefont {T.}~\bibnamefont {Liu}},\ and\ \bibinfo {author}
		{\bibfnamefont {S.}~\bibnamefont {Xiao}},\ }\bibfield  {title} {\bibinfo
		{title} {Symmetry-protected bound states in the continuum supported by
			all-dielectric metasurfaces},\ }\href
	{https://doi.org/10.1103/physreva.100.063803} {\bibfield  {journal} {\bibinfo
			{journal} {Phys. Rev. A}\ }\textbf {\bibinfo {volume} {100}},\ \bibinfo
		{pages} {063803} (\bibinfo {year} {2019})}\BibitemShut {NoStop}%
	\bibitem [{\citenamefont {Wang}\ \emph {et~al.}(2020)\citenamefont {Wang},
		\citenamefont {Duan}, \citenamefont {Chen}, \citenamefont {Zhou},
		\citenamefont {Liu},\ and\ \citenamefont {Xiao}}]{Wang2020}%
	\BibitemOpen
	\bibfield  {author} {\bibinfo {author} {\bibfnamefont {X.}~\bibnamefont
			{Wang}}, \bibinfo {author} {\bibfnamefont {J.}~\bibnamefont {Duan}}, \bibinfo
		{author} {\bibfnamefont {W.}~\bibnamefont {Chen}}, \bibinfo {author}
		{\bibfnamefont {C.}~\bibnamefont {Zhou}}, \bibinfo {author} {\bibfnamefont
			{T.}~\bibnamefont {Liu}},\ and\ \bibinfo {author} {\bibfnamefont
			{S.}~\bibnamefont {Xiao}},\ }\bibfield  {title} {\bibinfo {title}
		{Controlling light absorption of graphene at critical coupling through
			magnetic dipole quasi-bound states in the continuum resonance},\ }\href
	{https://doi.org/10.1103/physrevb.102.155432} {\bibfield  {journal} {\bibinfo
			{journal} {Phys. Rev. B}\ }\textbf {\bibinfo {volume} {102}},\ \bibinfo
		{pages} {155432} (\bibinfo {year} {2020})}\BibitemShut {NoStop}%
	\bibitem [{\citenamefont {Xiao}\ \emph {et~al.}(2022)\citenamefont {Xiao},
		\citenamefont {Qin}, \citenamefont {Duan}, \citenamefont {Wu},\ and\
		\citenamefont {Liu}}]{Xiao2022}%
	\BibitemOpen
	\bibfield  {author} {\bibinfo {author} {\bibfnamefont {S.}~\bibnamefont
			{Xiao}}, \bibinfo {author} {\bibfnamefont {M.}~\bibnamefont {Qin}}, \bibinfo
		{author} {\bibfnamefont {J.}~\bibnamefont {Duan}}, \bibinfo {author}
		{\bibfnamefont {F.}~\bibnamefont {Wu}},\ and\ \bibinfo {author}
		{\bibfnamefont {T.}~\bibnamefont {Liu}},\ }\bibfield  {title} {\bibinfo
		{title} {Polarization-controlled dynamically switchable high-harmonic
			generation from all-dielectric metasurfaces governed by dual bound states in
			the continuum},\ }\href {https://doi.org/10.1103/physrevb.105.195440}
	{\bibfield  {journal} {\bibinfo  {journal} {Phys. Rev. B}\ }\textbf {\bibinfo
			{volume} {105}},\ \bibinfo {pages} {195440} (\bibinfo {year}
		{2022})}\BibitemShut {NoStop}%
	\bibitem [{\citenamefont {Lu}\ \emph {et~al.}(2021)\citenamefont {Lu},
		\citenamefont {Dong}, \citenamefont {Tijiptoharsono}, \citenamefont {Ng},
		\citenamefont {Wang}, \citenamefont {Rezaei}, \citenamefont {Wang},
		\citenamefont {Leong}, \citenamefont {Lim}, \citenamefont {Yang} \emph
		{et~al.}}]{lu2021reversible}%
	\BibitemOpen
	\bibfield  {author} {\bibinfo {author} {\bibfnamefont {L.}~\bibnamefont
			{Lu}}, \bibinfo {author} {\bibfnamefont {Z.}~\bibnamefont {Dong}}, \bibinfo
		{author} {\bibfnamefont {F.}~\bibnamefont {Tijiptoharsono}}, \bibinfo
		{author} {\bibfnamefont {R.~J.~H.}\ \bibnamefont {Ng}}, \bibinfo {author}
		{\bibfnamefont {H.}~\bibnamefont {Wang}}, \bibinfo {author} {\bibfnamefont
			{S.~D.}\ \bibnamefont {Rezaei}}, \bibinfo {author} {\bibfnamefont
			{Y.}~\bibnamefont {Wang}}, \bibinfo {author} {\bibfnamefont {H.~S.}\
			\bibnamefont {Leong}}, \bibinfo {author} {\bibfnamefont {P.~C.}\ \bibnamefont
			{Lim}}, \bibinfo {author} {\bibfnamefont {J.~K.}\ \bibnamefont {Yang}}, \emph
		{et~al.},\ }\bibfield  {title} {\bibinfo {title} {Reversible tuning of mie
			resonances in the visible spectrum},\ }\href
	{https://doi.org/10.1021/acsnano.1c07114} {\bibfield  {journal} {\bibinfo
			{journal} {ACS Nano}\ }\textbf {\bibinfo {volume} {15}},\ \bibinfo {pages}
		{19722} (\bibinfo {year} {2021})}\BibitemShut {NoStop}%
	\bibitem [{\citenamefont {Lu}\ \emph {et~al.}(2022)\citenamefont {Lu},
		\citenamefont {Reniers}, \citenamefont {Wang}, \citenamefont {Jiao},\ and\
		\citenamefont {Simpson}}]{lu2022reconfigurable}%
	\BibitemOpen
	\bibfield  {author} {\bibinfo {author} {\bibfnamefont {L.}~\bibnamefont
			{Lu}}, \bibinfo {author} {\bibfnamefont {S.}~\bibnamefont {Reniers}},
		\bibinfo {author} {\bibfnamefont {Y.}~\bibnamefont {Wang}}, \bibinfo {author}
		{\bibfnamefont {Y.}~\bibnamefont {Jiao}},\ and\ \bibinfo {author}
		{\bibfnamefont {R.~E.}\ \bibnamefont {Simpson}},\ }\bibfield  {title}
	{\bibinfo {title} {Reconfigurable inp waveguide components using the sb2s3
			phase change material},\ }\href {https://doi.org/10.1088/2040-8986/ac7e5a}
	{\bibfield  {journal} {\bibinfo  {journal} {J. Opt.}\ }\textbf {\bibinfo
			{volume} {9}},\ \bibinfo {pages} {094001} (\bibinfo {year}
		{2022})}\BibitemShut {NoStop}%
	\bibitem [{\citenamefont {Moitra}\ \emph {et~al.}(2022)\citenamefont {Moitra},
		\citenamefont {Wang}, \citenamefont {Liang}, \citenamefont {Lu},
		\citenamefont {Poh}, \citenamefont {Mass}, \citenamefont {Simpson},
		\citenamefont {Kuznetsov},\ and\ \citenamefont
		{Paniagua-Dominguez}}]{Moitra2022}%
	\BibitemOpen
	\bibfield  {author} {\bibinfo {author} {\bibfnamefont {P.}~\bibnamefont
			{Moitra}}, \bibinfo {author} {\bibfnamefont {Y.}~\bibnamefont {Wang}},
		\bibinfo {author} {\bibfnamefont {X.}~\bibnamefont {Liang}}, \bibinfo
		{author} {\bibfnamefont {L.}~\bibnamefont {Lu}}, \bibinfo {author}
		{\bibfnamefont {A.}~\bibnamefont {Poh}}, \bibinfo {author} {\bibfnamefont
			{T.~W.}\ \bibnamefont {Mass}}, \bibinfo {author} {\bibfnamefont {R.~E.}\
			\bibnamefont {Simpson}}, \bibinfo {author} {\bibfnamefont {A.~I.}\
			\bibnamefont {Kuznetsov}},\ and\ \bibinfo {author} {\bibfnamefont
			{R.}~\bibnamefont {Paniagua-Dominguez}},\ }\bibfield  {title} {\bibinfo
		{title} {Programmable wavefront control in the visible spectrum using
			low-loss chalcogenide phase-change metasurfaces},\ }\href
	{https://doi.org/10.1002/adma.202205367} {\bibfield  {journal} {\bibinfo
			{journal} {Adv. Mater.}\ }\textbf {\bibinfo {volume} {35}},\ \bibinfo {pages}
		{2205367} (\bibinfo {year} {2022})}\BibitemShut {NoStop}%
	\bibitem [{\citenamefont {Tian}\ \emph {et~al.}(2019)\citenamefont {Tian},
		\citenamefont {Luo}, \citenamefont {Yang}, \citenamefont {Ding},
		\citenamefont {Qu}, \citenamefont {Zhao}, \citenamefont {Qiu},\ and\
		\citenamefont {Bozhevolnyi}}]{Tian2019}%
	\BibitemOpen
	\bibfield  {author} {\bibinfo {author} {\bibfnamefont {J.}~\bibnamefont
			{Tian}}, \bibinfo {author} {\bibfnamefont {H.}~\bibnamefont {Luo}}, \bibinfo
		{author} {\bibfnamefont {Y.}~\bibnamefont {Yang}}, \bibinfo {author}
		{\bibfnamefont {F.}~\bibnamefont {Ding}}, \bibinfo {author} {\bibfnamefont
			{Y.}~\bibnamefont {Qu}}, \bibinfo {author} {\bibfnamefont {D.}~\bibnamefont
			{Zhao}}, \bibinfo {author} {\bibfnamefont {M.}~\bibnamefont {Qiu}},\ and\
		\bibinfo {author} {\bibfnamefont {S.~I.}\ \bibnamefont {Bozhevolnyi}},\
	}\bibfield  {title} {\bibinfo {title} {Active control of anapole states by
			structuring the phase-change alloy ge2sb2te5},\ }\href
	{https://doi.org/10.1038/s41467-018-08057-1} {\bibfield  {journal} {\bibinfo
			{journal} {Nat. Commun.}\ }\textbf {\bibinfo {volume} {10}},\ \bibinfo
		{pages} {396} (\bibinfo {year} {2019})}\BibitemShut {NoStop}%
	\bibitem [{\citenamefont {Meng}\ \emph {et~al.}(2021)\citenamefont {Meng},
		\citenamefont {Chen}, \citenamefont {Xu}, \citenamefont {Zhu}, \citenamefont
		{Yuan},\ and\ \citenamefont {Zhang}}]{meng2021high}%
	\BibitemOpen
	\bibfield  {author} {\bibinfo {author} {\bibfnamefont {Q.}~\bibnamefont
			{Meng}}, \bibinfo {author} {\bibfnamefont {X.}~\bibnamefont {Chen}}, \bibinfo
		{author} {\bibfnamefont {W.}~\bibnamefont {Xu}}, \bibinfo {author}
		{\bibfnamefont {Z.}~\bibnamefont {Zhu}}, \bibinfo {author} {\bibfnamefont
			{X.}~\bibnamefont {Yuan}},\ and\ \bibinfo {author} {\bibfnamefont
			{J.}~\bibnamefont {Zhang}},\ }\bibfield  {title} {\bibinfo {title} {High q
			resonant sb2s3-lithium niobate metasurface for active nanophotonics},\
	}\href@noop {} {\bibfield  {journal} {\bibinfo  {journal} {Nanomaterials}\
		}\textbf {\bibinfo {volume} {11}},\ \bibinfo {pages} {2373} (\bibinfo {year}
		{2021})}\BibitemShut {NoStop}%
	\bibitem [{\citenamefont {Liu}\ \emph {et~al.}(2022)\citenamefont {Liu},
		\citenamefont {Han}, \citenamefont {Duan},\ and\ \citenamefont
		{Xiao}}]{Liu2022}%
	\BibitemOpen
	\bibfield  {author} {\bibinfo {author} {\bibfnamefont {T.}~\bibnamefont
			{Liu}}, \bibinfo {author} {\bibfnamefont {Z.}~\bibnamefont {Han}}, \bibinfo
		{author} {\bibfnamefont {J.}~\bibnamefont {Duan}},\ and\ \bibinfo {author}
		{\bibfnamefont {S.}~\bibnamefont {Xiao}},\ }\bibfield  {title} {\bibinfo
		{title} {Phase-change metasurfaces for dynamic image display and information
			encryption},\ }\href {https://doi.org/10.1103/physrevapplied.18.044078}
	{\bibfield  {journal} {\bibinfo  {journal} {Phys. Rev. Appl.}\ }\textbf
		{\bibinfo {volume} {18}},\ \bibinfo {pages} {044078} (\bibinfo {year}
		{2022})}\BibitemShut {NoStop}%
	\bibitem [{\citenamefont {Lee}\ \emph {et~al.}(2017)\citenamefont {Lee},
		\citenamefont {Kim}, \citenamefont {Cho}, \citenamefont {Kim}, \citenamefont
		{Kim}, \citenamefont {Ryu}, \citenamefont {Kim}, \citenamefont {Kang},
		\citenamefont {Hwang},\ and\ \citenamefont {Hwang}}]{Lee2017}%
	\BibitemOpen
	\bibfield  {author} {\bibinfo {author} {\bibfnamefont {S.-Y.}\ \bibnamefont
			{Lee}}, \bibinfo {author} {\bibfnamefont {Y.-H.}\ \bibnamefont {Kim}},
		\bibinfo {author} {\bibfnamefont {S.-M.}\ \bibnamefont {Cho}}, \bibinfo
		{author} {\bibfnamefont {G.~H.}\ \bibnamefont {Kim}}, \bibinfo {author}
		{\bibfnamefont {T.-Y.}\ \bibnamefont {Kim}}, \bibinfo {author} {\bibfnamefont
			{H.}~\bibnamefont {Ryu}}, \bibinfo {author} {\bibfnamefont {H.~N.}\
			\bibnamefont {Kim}}, \bibinfo {author} {\bibfnamefont {H.~B.}\ \bibnamefont
			{Kang}}, \bibinfo {author} {\bibfnamefont {C.-Y.}\ \bibnamefont {Hwang}},\
		and\ \bibinfo {author} {\bibfnamefont {C.-S.}\ \bibnamefont {Hwang}},\
	}\bibfield  {title} {\bibinfo {title} {Holographic image generation with a
			thin-film resonance caused by chalcogenide phase-change material},\ }\href
	{https://doi.org/10.1038/srep41152} {\bibfield  {journal} {\bibinfo
			{journal} {Sci. Rep.}\ }\textbf {\bibinfo {volume} {7}},\ \bibinfo {pages}
		{41152} (\bibinfo {year} {2017})}\BibitemShut {NoStop}%
	\bibitem [{\citenamefont {Zhou}\ \emph {et~al.}(2020)\citenamefont {Zhou},
		\citenamefont {Wang}, \citenamefont {Li}, \citenamefont {Wang}, \citenamefont
		{Wei}, \citenamefont {Geng},\ and\ \citenamefont {Huang}}]{Zhou2020}%
	\BibitemOpen
	\bibfield  {author} {\bibinfo {author} {\bibfnamefont {H.}~\bibnamefont
			{Zhou}}, \bibinfo {author} {\bibfnamefont {Y.}~\bibnamefont {Wang}}, \bibinfo
		{author} {\bibfnamefont {X.}~\bibnamefont {Li}}, \bibinfo {author}
		{\bibfnamefont {Q.}~\bibnamefont {Wang}}, \bibinfo {author} {\bibfnamefont
			{Q.}~\bibnamefont {Wei}}, \bibinfo {author} {\bibfnamefont {G.}~\bibnamefont
			{Geng}},\ and\ \bibinfo {author} {\bibfnamefont {L.}~\bibnamefont {Huang}},\
	}\bibfield  {title} {\bibinfo {title} {Switchable active phase modulation and
			holography encryption based on hybrid metasurfaces},\ }\href
	{https://doi.org/10.1515/nanoph-2019-0519} {\bibfield  {journal} {\bibinfo
			{journal} {Nanophotonics}\ }\textbf {\bibinfo {volume} {9}},\ \bibinfo
		{pages} {905} (\bibinfo {year} {2020})}\BibitemShut {NoStop}%
	\bibitem [{\citenamefont {Mao}\ \emph {et~al.}(2020)\citenamefont {Mao},
		\citenamefont {Li}, \citenamefont {Li}, \citenamefont {Zhang},\ and\
		\citenamefont {Cao}}]{Mao2020}%
	\BibitemOpen
	\bibfield  {author} {\bibinfo {author} {\bibfnamefont {L.}~\bibnamefont
			{Mao}}, \bibinfo {author} {\bibfnamefont {Y.}~\bibnamefont {Li}}, \bibinfo
		{author} {\bibfnamefont {G.}~\bibnamefont {Li}}, \bibinfo {author}
		{\bibfnamefont {S.}~\bibnamefont {Zhang}},\ and\ \bibinfo {author}
		{\bibfnamefont {T.}~\bibnamefont {Cao}},\ }\bibfield  {title} {\bibinfo
		{title} {Reversible switching of electromagnetically induced transparency in
			phase change metasurfaces},\ }\href {https://doi.org/10.1117/1.ap.2.5.056004}
	{\bibfield  {journal} {\bibinfo  {journal} {Adv. Photonics}\ }\textbf
		{\bibinfo {volume} {2}},\ \bibinfo {pages} {056004} (\bibinfo {year}
		{2020})}\BibitemShut {NoStop}%
	\bibitem [{\citenamefont {Liu}\ \emph {et~al.}(2021)\citenamefont {Liu},
		\citenamefont {Fang},\ and\ \citenamefont {Xiao}}]{Liu2021}%
	\BibitemOpen
	\bibfield  {author} {\bibinfo {author} {\bibfnamefont {T.}~\bibnamefont
			{Liu}}, \bibinfo {author} {\bibfnamefont {X.}~\bibnamefont {Fang}},\ and\
		\bibinfo {author} {\bibfnamefont {S.}~\bibnamefont {Xiao}},\ }\bibfield
	{title} {\bibinfo {title} {Tuning nonlinear second-harmonic generation in
			{AlGaAs} nanoantennas via chalcogenide phase-change material},\ }\href
	{https://doi.org/10.1103/physrevb.104.195428} {\bibfield  {journal} {\bibinfo
			{journal} {Phys. Rev. B}\ }\textbf {\bibinfo {volume} {104}},\ \bibinfo
		{pages} {195428} (\bibinfo {year} {2021})}\BibitemShut {NoStop}%
	\bibitem [{\citenamefont {Chen}\ \emph {et~al.}(2012)\citenamefont {Chen},
		\citenamefont {Huang}, \citenamefont {Mühlenbernd}, \citenamefont {Li},
		\citenamefont {Bai}, \citenamefont {Tan}, \citenamefont {Jin}, \citenamefont
		{Qiu}, \citenamefont {Zhang},\ and\ \citenamefont {Zentgraf}}]{Chen2012}%
	\BibitemOpen
	\bibfield  {author} {\bibinfo {author} {\bibfnamefont {X.}~\bibnamefont
			{Chen}}, \bibinfo {author} {\bibfnamefont {L.}~\bibnamefont {Huang}},
		\bibinfo {author} {\bibfnamefont {H.}~\bibnamefont {Mühlenbernd}}, \bibinfo
		{author} {\bibfnamefont {G.}~\bibnamefont {Li}}, \bibinfo {author}
		{\bibfnamefont {B.}~\bibnamefont {Bai}}, \bibinfo {author} {\bibfnamefont
			{Q.}~\bibnamefont {Tan}}, \bibinfo {author} {\bibfnamefont {G.}~\bibnamefont
			{Jin}}, \bibinfo {author} {\bibfnamefont {C.-W.}\ \bibnamefont {Qiu}},
		\bibinfo {author} {\bibfnamefont {S.}~\bibnamefont {Zhang}},\ and\ \bibinfo
		{author} {\bibfnamefont {T.}~\bibnamefont {Zentgraf}},\ }\bibfield  {title}
	{\bibinfo {title} {Dual-polarity plasmonic metalens for visible light},\
	}\href {https://doi.org/10.1038/ncomms2207} {\bibfield  {journal} {\bibinfo
			{journal} {Nat. Commun.}\ }\textbf {\bibinfo {volume} {3}},\ \bibinfo {pages}
		{1198} (\bibinfo {year} {2012})}\BibitemShut {NoStop}%
	\bibitem [{\citenamefont {Khorasaninejad}\ \emph {et~al.}(2016)\citenamefont
		{Khorasaninejad}, \citenamefont {Chen}, \citenamefont {Devlin}, \citenamefont
		{Oh}, \citenamefont {Zhu},\ and\ \citenamefont
		{Capasso}}]{Khorasaninejad2016}%
	\BibitemOpen
	\bibfield  {author} {\bibinfo {author} {\bibfnamefont {M.}~\bibnamefont
			{Khorasaninejad}}, \bibinfo {author} {\bibfnamefont {W.~T.}\ \bibnamefont
			{Chen}}, \bibinfo {author} {\bibfnamefont {R.~C.}\ \bibnamefont {Devlin}},
		\bibinfo {author} {\bibfnamefont {J.}~\bibnamefont {Oh}}, \bibinfo {author}
		{\bibfnamefont {A.~Y.}\ \bibnamefont {Zhu}},\ and\ \bibinfo {author}
		{\bibfnamefont {F.}~\bibnamefont {Capasso}},\ }\bibfield  {title} {\bibinfo
		{title} {Metalenses at visible wavelengths: Diffraction-limited focusing and
			subwavelength resolution imaging},\ }\href
	{https://doi.org/10.1126/science.aaf6644} {\bibfield  {journal} {\bibinfo
			{journal} {Science}\ }\textbf {\bibinfo {volume} {352}},\ \bibinfo {pages}
		{1190} (\bibinfo {year} {2016})}\BibitemShut {NoStop}%
	\bibitem [{\citenamefont {Zhao}\ \emph {et~al.}(2021)\citenamefont {Zhao},
		\citenamefont {Chen}, \citenamefont {Zhuang}, \citenamefont {Zhang},
		\citenamefont {Chen}, \citenamefont {Chen}, \citenamefont {Liu},
		\citenamefont {Wang}, \citenamefont {Chen}, \citenamefont {Wang},
		\citenamefont {Liu}, \citenamefont {Yin}, \citenamefont {Xiao}, \citenamefont
		{Shi}, \citenamefont {Dong}, \citenamefont {Zi},\ and\ \citenamefont
		{Tsai}}]{Zhao2021}%
	\BibitemOpen
	\bibfield  {author} {\bibinfo {author} {\bibfnamefont {M.}~\bibnamefont
			{Zhao}}, \bibinfo {author} {\bibfnamefont {M.~K.}\ \bibnamefont {Chen}},
		\bibinfo {author} {\bibfnamefont {Z.-P.}\ \bibnamefont {Zhuang}}, \bibinfo
		{author} {\bibfnamefont {Y.}~\bibnamefont {Zhang}}, \bibinfo {author}
		{\bibfnamefont {A.}~\bibnamefont {Chen}}, \bibinfo {author} {\bibfnamefont
			{Q.}~\bibnamefont {Chen}}, \bibinfo {author} {\bibfnamefont {W.}~\bibnamefont
			{Liu}}, \bibinfo {author} {\bibfnamefont {J.}~\bibnamefont {Wang}}, \bibinfo
		{author} {\bibfnamefont {Z.-M.}\ \bibnamefont {Chen}}, \bibinfo {author}
		{\bibfnamefont {B.}~\bibnamefont {Wang}}, \bibinfo {author} {\bibfnamefont
			{X.}~\bibnamefont {Liu}}, \bibinfo {author} {\bibfnamefont {H.}~\bibnamefont
			{Yin}}, \bibinfo {author} {\bibfnamefont {S.}~\bibnamefont {Xiao}}, \bibinfo
		{author} {\bibfnamefont {L.}~\bibnamefont {Shi}}, \bibinfo {author}
		{\bibfnamefont {J.-W.}\ \bibnamefont {Dong}}, \bibinfo {author}
		{\bibfnamefont {J.}~\bibnamefont {Zi}},\ and\ \bibinfo {author}
		{\bibfnamefont {D.~P.}\ \bibnamefont {Tsai}},\ }\bibfield  {title} {\bibinfo
		{title} {Phase characterisation of metalenses},\ }\href
	{https://doi.org/10.1038/s41377-021-00492-y} {\bibfield  {journal} {\bibinfo
			{journal} {Light Sci. Appl.}\ }\textbf {\bibinfo {volume} {10}},\ \bibinfo
		{pages} {52} (\bibinfo {year} {2021})}\BibitemShut {NoStop}%
	\bibitem [{\citenamefont {Li}\ \emph {et~al.}(2022)\citenamefont {Li},
		\citenamefont {Li}, \citenamefont {Yue}, \citenamefont {Zheng}, \citenamefont
		{Wang}, \citenamefont {Liu}, \citenamefont {Xu}, \citenamefont {Song},
		\citenamefont {Yang}, \citenamefont {Li}, \citenamefont {Li}, \citenamefont
		{Tang}, \citenamefont {Zhang}, \citenamefont {Zhang},\ and\ \citenamefont
		{Yao}}]{Li2022}%
	\BibitemOpen
	\bibfield  {author} {\bibinfo {author} {\bibfnamefont {J.}~\bibnamefont
			{Li}}, \bibinfo {author} {\bibfnamefont {J.}~\bibnamefont {Li}}, \bibinfo
		{author} {\bibfnamefont {Z.}~\bibnamefont {Yue}}, \bibinfo {author}
		{\bibfnamefont {C.}~\bibnamefont {Zheng}}, \bibinfo {author} {\bibfnamefont
			{G.}~\bibnamefont {Wang}}, \bibinfo {author} {\bibfnamefont {J.}~\bibnamefont
			{Liu}}, \bibinfo {author} {\bibfnamefont {H.}~\bibnamefont {Xu}}, \bibinfo
		{author} {\bibfnamefont {C.}~\bibnamefont {Song}}, \bibinfo {author}
		{\bibfnamefont {F.}~\bibnamefont {Yang}}, \bibinfo {author} {\bibfnamefont
			{H.}~\bibnamefont {Li}}, \bibinfo {author} {\bibfnamefont {F.}~\bibnamefont
			{Li}}, \bibinfo {author} {\bibfnamefont {T.}~\bibnamefont {Tang}}, \bibinfo
		{author} {\bibfnamefont {Y.}~\bibnamefont {Zhang}}, \bibinfo {author}
		{\bibfnamefont {Y.}~\bibnamefont {Zhang}},\ and\ \bibinfo {author}
		{\bibfnamefont {J.}~\bibnamefont {Yao}},\ }\bibfield  {title} {\bibinfo
		{title} {Structured vector field manipulation of terahertz wave along the
			propagation direction based on dielectric metasurfaces},\ }\href
	{https://doi.org/10.1002/lpor.202200325} {\bibfield  {journal} {\bibinfo
			{journal} {Laser Photonics Rev.}\ }\textbf {\bibinfo {volume} {16}},\
		\bibinfo {pages} {2200325} (\bibinfo {year} {2022})}\BibitemShut {NoStop}%
\end{thebibliography}
%

\end{document}